\documentstyle[12pt]{article}
 
\begin{document}
\def\omt{\tilde{\omega}}
\def\ti{\tilde}
\def\o{\Omega}
\def\bchi{\bar\chi^i}
\def\In{{\rm Int}}
\def\ba{\bar a}
\def\w{\wedge}
\def\ep{\epsilon}
\def\k{\kappa}
\def\Tr{{\rm Tr}}
\def\ST{{\rm STr}}
\def\ss{\subset}
\def\rn{\vert \alpha\vert^2}
\def\bi{\bibitem}
\def\ot{\oti\def\om{\omega}
\dmes}
\def\bc{{\bf C}}
\def\bz{{\bf Z}}
\def\ptp{\stackrel{\otimes}{,}}
\def\br{{\bf R}}
\def\de{\delta}
 \def\bt{\beta}
 \def\ve{\vert}
\def\al{\alpha}
\def\la{\langle}
\def\ra{\rangle}
\def\Ga{\Gamma}
\def\st{\stackrel{\wedge}{,}}
\def\stv{\stackrel{\wedge}{\vert}}
\def\th{\theta}
\def\lm{\ti\lambda}
\def\U{\Upsilon}
\def\jp{{1\over 2}}
\def\js{{1\over 4}}
\def\d{\partial}
\def\tr{\triangleright}
\def\trl{\triangleleft}
\def\d{\partial}
\def\bq{\}_{P}}
\def\be{\begin{equation}}
\def\ee{\end{equation}}
\def\bea{\begin{eqnarray}}
\def\eea{\end{eqnarray}}
\def\D{{\cal D}}
\def\A{{\cal A}}
\def\G{{\cal G}}
\def\K{{\cal K}}
\def\H{{\cal H}}
\def\P{{\cal P}}
\def\N{{\cal N}}
\def\R{{\cal R}}
\def\B{{\cal B}}
\def\T{{\cal T}}
\def\bT{\bar{\cal T}}
\def\F{{\cal F}}
\def\n{{1\over n}}
\def\si{\sigma}
\def\ta{\tau}
\def\ov{\over}
\def\l{\lambda}
\def\L{\Lambda}
\def\lpb{{\bf \{}}
\def\rpb{{\bf \}}}

\def\pih{\hat{\pi}}
 
\def\U{\Upsilon}
\def\e{\varepsilon}
\def\bt{\beta}
\def\ga{\gamma}
\def\om{\omega}
\def\be{\begin{equation}}
\def\ee{\end{equation}}
\def\bea{\begin{eqnarray}}
\def\eea{\end{eqnarray}}
\def\D{{\cal D}}
\def\C{{\cal C}}
\def\G{{\cal G}}
\def\H{{\cal H}}
\def\R{{\cal R}}
\def\B{{\cal B}}
\def\K{{\cal K}}
\def\T{{\cal T}}
\def\S{{\cal S}}
\def\bT{\bar{\cal T}}
\def\F{{\cal F}}
\def\n{{1\over n}}
\def\si{\sigma}
\def\ta{\tau}
\def\ot{\otimes}
\def\l{\lambda}
\def\L{\Lambda}
\def\ve{\vert}
\def\nr{\nabla_R}
\def\nl{\nabla_L}
\def\pih{\hat{\pi}}
 
\def\e{\varepsilon}
\def\bt{\beta}
\def\ga{\gamma}

\sloppy \raggedbottom
\setcounter{page}{1}

\newpage
\setcounter{figure}{0}
\setcounter{equation}{0}
\setcounter{footnote}{0}
\setcounter{table}{0}
\setcounter{section}{0}

\begin{titlepage}
\begin{flushright}
{}~
IML 2006-05\\
math-ph/0602048
\end{flushright}

\vspace{3cm}
\begin{center}
{\Large \bf  On moment maps associated to a twisted Heisenberg double}\\ 
[50pt]{\small
{\bf C. Klim\v{c}\'{\i}k }
\\ ~~\\Institute de math\'ematiques de Luminy,
 \\163, Avenue de Luminy, 13288 Marseille, France}

\vspace{1cm}

\begin{abstract}
We review the concept of the (anomalous) Poisson-Lie symmetry in a way that emphasises
the notion of Poisson-Lie Hamiltonian. The  language that we develop  turns out to be very useful
for several applications:   we prove that 
 the   left and the  right actions of a group $G$ on its twisted Heisenberg double $(D,\kappa)$ realize the (anomalous) Poisson-Lie symmetries  and  we explain  in a very transparent way the concept of the Poisson-Lie subsymmetry and  that of Poisson-Lie symplectic reduction. Under some additional conditions, we construct   also  a non-anomalous   moment map corresponding to a sort of quasi-adjoint action of $G$ on  $(D,\kappa)$.  The absence of the anomaly of this "quasi-adjoint" moment map permits
 to perform the   gauging of deformed  WZW models.  
      \end{abstract}
\vspace{1cm}

\end{center}
\end{titlepage}
\newpage

\section{Introduction} Poisson-Lie symmetry \cite{ST1} is the generalization of the ordinary Hamiltonian symmetry of a dynamical system and, upon quantizing, it becomes the quantum group symmetry.
Many dynamical systems can be deformed in such a way that  their  ordinary symmetries become
Poisson-Lie. Among such systems there is also the standard WZW model \cite{W} where the loop group symmetry gets deformed \cite{K}.  The principal goal of the present work is to develop the theory
of gauging of the deformed WZW model.   

\medskip

\noindent  From the mathematical point of view, the problem amounts
to identify non-anomalous Poisson-Lie subsymmetries of the deformed WZW model  which would permit to perform the gauging.   In order to describe the Poisson-Lie analogue of the WZW vanishing anomaly condition \cite{Wi3},   first we shall have  to develop   appropriate mathematical tools.  It particular, it turns out that the standard definition of the Poisson-Lie
symmetry (i.e.   the action map $G\times M\to M$ is Poisson)  is too rough since it is unable to distinguish between non-anomalous and anomalous symmetries. For this reason, we shall
refine the standard concept of the Poisson-Lie symmetry and propose its new definition based 
rather on the Poisson-Lie structure on the cosymmetry (or dual) group $B$ than on the symmetry  group $G$.   We are fully aware that the language that we develop is not quite standard
 in the Poisson-(Lie) geometry  but we find it   well adapted for our discussion of anomalies  and we also believe that  it may constitute an insightful alternative in treating the Poisson-Lie symmetric 
 systems in general. 
 
 \medskip

\noindent  The central object of our investigations will be  a class of   Poisson manifolds introduced by Semenov-Tian-Shansky  under the name  of twisted Heisenberg doubles  \cite{ST2}.     As it was conjectured in \cite{K} and showed  in \cite{K2}, particular elements of this class play the role of the phase spaces of the deformed WZW models.    This also means that results obtained in full generality for any  twisted Heisenberg
double will also hold for any deformed WZW model.

 \medskip

\noindent  In order to present in this introduction the  principal  ideas and results  of our work, we first   expose two main definitions and three main theorems   proved later in the body of the  paper.

  \vskip1pc
 
\noindent \underline{Definition 1}:  Let $M$ be a   symplectic  manifold whose algebra of smooth functions $Fun(M)$ is 
equipped with a Poisson bracket $\{.,.\}$. Let $B$ be  a Poisson-Lie group and let $\mu: M\to B$ be a smooth   map.
 To every function $y\in Fun(B)$ we can associate a vector field $w_\mu(y)\in Vect(M)$ as follows:   
  $$w_\mu(y)f=\{f,\mu^*(y')\} \mu^*(S(y'')), \quad y\in Fun(B), f\in Fun(M).$$ 
We say that   $\mu$ realizes the Poisson-Lie symmetry of   $M$ if  the map 
$w_\mu$ is  homomorphism of the  Lie algebras  $Fun(B)$ and $Vect(M)$.  If, moreover, the map $\mu$
is Poisson, we say that the symmetry is equivariant or non-anomalous.

\vskip1pc

\noindent \underline{Definition 2}:  Let $D$ be an even-dimensional Lie group equipped with a maximally Lorentzian
bi-invariant metric. If  
$Lie(D)=Lie(G)\stackrel{.}{+}Lie(B)$, where 
$G$ and $B$ are maximally isotropic  subgroups, $D$ is called the Drinfeld double of $G$ or the Drinfeld double of $B$.  Let $\k$ be a  metric preserving 
automorphism of $D$   and suppose that there are  respective basis  $T^i$ and $t_i$ 
($i=1,...,n$)  of $\G=Lie(G)$ and $\B=Lie(B)$ such that 
 $$(T^i,t_j)_\D=\delta^i_j.$$
Then the (basis independent) expression 
$$\{f_1,f_2\}_D\equiv \nabla^R_{T^i}f_1 \nabla^R_{t_i}f_2 -
\nabla^L_{\k(t_i)}f_1\nabla^L_{\k(T^i)}f_2, \quad f_1,f_2\in Fun(D) $$
is a  Poisson bracket and the Poisson manifold $(D,\{.,.\}_D)$ is called the twisted Heisenberg double.

 \vskip1pc

\noindent \underline{Theorem 1}:
\noindent Let $D$ be a  twisted Heisenberg double which is also decomposable, i.e. such that two global unambiguous decompositions hold: $D=\k(B)G$ 
and $D=\k(G)B$.  Consider  (smooth)  maps $\L_{L}, \L_R: D\to B$, $\Xi_R,\Xi_L:D\to G$   respectively induced  by these two decompositions. Then  it holds:

\medskip

\noindent a)  The Poisson manifold $(D,\{.,.\}_D)$ is   symplectic. 
 
  \noindent b) Both 
maps $\L_L$ and $\L_R$ realize the (anomalous)  Poisson-Lie symmetries of the symplectic manifold 
$(D,\{.,.\}_D)$.  The corresponding symmetry group is $G$ acting  as 
$$h\tr K=\k(h)K , \quad h\in G,\quad  K\in D$$
or, respectively, as
 $$h\tr K=Kh^{-1} , \quad h\in G,\quad  K\in D.$$
\vskip1pc

\noindent \underline{Theorem 2}:
\noindent Let $D$ be a  decomposable twisted Heisenberg double such that  the twisting automorphism $\k$ 
preserves the subgroup $B$.   Construct two new maps $B_L:D\to B$
 and $B_R:D\to B$ as follows
 $$B_L(K)=\k(\L_L(K))\L_R(K),\quad B_R(K)=\k^{-1}(\L_R(K))\L_L(K), \quad K\in D.$$
 Then it holds:  Both  maps $B_L$ and $B_R$ are Poisson and they  realize the  (non-anomalous) Poisson-Lie 
symmetries of $(D,\{.,.\}_D)$. The corresponding symmetry group is $G$ acting  as 
$$h\tr K = \k(h)K\Xi_R(\k[h\L_L(K)]), \quad h\in G,\ K\in D,$$
or, respectively, as
$$h\tr K =\k[\Xi_L^{-1}(\L_R^{-1}(K)h^{-1})]Kh^{-1}.\quad h\in G, \ K\in D.$$

 \vskip1pc
 
\noindent \underline{Theorem 3}:  Let $D$ be a  decomposable twisted Heisenberg double, $\kappa$ an automorphism
of $D$ preserving  $B$ and $N$   a normal subgroup  of $B$. Denote by  $C$  the   factor group $B/N$,  by    $\rho$   the natural homomorphism $ B\to C$ and by $P_\k:Lie(D)\to Lie(B)$ a projector on $Lie(B)$ with kernel
$\k(Lie(G))$.    Suppose that the  Hopf subalgebra  $\rho^*(Fun(C))$ of $Fun(B)$  is also a Poisson subalgebra.    
 Then it holds:   The  composed map
 $\nu_R\equiv \rho\circ \L_R$ realizes  the Poisson-Lie symmetry of $D$ and the corresponding 
 symmetry group $H$ is   the subgroup of $G$.    If, moreover,  $P_\k(Lie(H))\subset Lie(N)$ then the moment map $\nu_R$ is non-anomalous.

\vskip1pc

\noindent  Apart from these three theorems, we prove   two more propositions (Lemma 3 and Lemma 4) enlarging the story to the
non-decomposable  twisted Heisenberg doubles. The  formulations of those additional Lemmas require introduction of several new concepts therefore, for the sake of conciseness of this introduction,  we shall expose them only in Section 3.3.
 
 \vskip 1pc
 
 \noindent The principal field of applications of our results is the theory of non-linear $\si$-models which are two-dimensional field theories describing the propagation of closed strings on a Riemannian manifold $T$. The manifold $T$ is often referred to as the target space and it comes also   equipped with a closed 3-form $H$. The classical action for a closed string
configuration $x^\mu(\si,\tau)$ reads 
 $$S[x^\mu(\si,\tau)]=\jp\int d\sigma d\tau G_{\mu\nu}(x) \partial_+ x^\mu \partial_- x^\nu +\int _V  x^*H,$$
where $\si$ is a periodic loop parameter, $\tau$ the evolution parameter, $x^\mu$ are coordinates on $T$, $G_{\mu\nu}$  are the
components  of the Riemannian metric and
$$\partial_\pm\equiv \partial_\tau\pm\partial_\si.$$
 It should be noted that the configuration $x^\mu(\si,\tau)$ is extended to a configuration  
 defined in the volume $V$ whose boundary is the surface of the propagating closed string and
 $x^*H$ is  the pull-back of the $H$-potential  to this volume $V$. A detailed explanation why the
 variational principle based on the action $S$ does not depend on the ambiguity of the extension of $ x$
 is given e.g. in \cite{W,Gaw2,KS}. The prominent example of the non-linear $\si$-model is the WZW model for which the target space is the
compact group manifold $K$ equipped with the standard Killing-Cartan metric $(.,.)_\K$. Its action reads
$$S_{WZW}[g(\si,\tau)]=\jp\int d\sigma d\tau (\partial_+gg^{-1},\partial_-gg^{-1})_\K +{1\over 12}\int_V ([dgg^{-1},dgg^{-1}], dgg^{-1})_\K.$$
Let $S$ be a subgroup of $K$ and let $A_\pm(\si,\tau)$ be two $Lie(S)$-valued fields.
The gauged $K/S$ WZW model is then a dynamical system described by the following classical action
$$S_{GWZW}[g(\si,\tau),A_\pm(\si,\tau)]=S_{WZW}[g(\si,\tau)] +$$
$$+\int d\sigma d\tau \biggl(-(\partial_+gg^{-1},A_-)_\K
+(\partial_-gg^{-1},A_+)_\K -(g^{-1}A_-g,A_+)_\K +(A_-,A_+)_\K\biggr).$$
The action $S_{GWZW}$ is invariant with respect to gauge transformations
$$g(\si,\tau)\to s^{-1}(\si,\tau)g(\si,\tau) s(\si,\tau), $$ $$A_\pm(\si,\tau)\to s^{-1}(\si,\tau)A_\pm(\si,\tau) s(\si,\tau)
-s^{-1}(\si,\tau)\partial_\pm s(\si,\tau),$$
where $s(\si,\tau)$ takes values in the subgroup $S$.

\medskip 
 
 \noindent (Gauged) WZW models are dynamical systems whose phase spaces are symplectic manifolds. 
  We shall show in Section 4, that their symplectic structures coincide with those  of (gauged)  twisted Heisenberg doubles.
  Actually, the twisted Heisenberg doubles underlying the ordinary WZW models are very special in the sense that the  symmetry
  group $G$ is the loop group $LK$ and the 
  cosymmetry group $B$ is Abelian.  If we consider also doubles with non-Abelian $B$, we are very naturally led to more
  general theories which we call the deformed WZW models.  Let us now explain the meaning of the Theorems 1,2 and 3 in the
  WZW context. 
  
  \medskip 
  
  \noindent If $B$ is Abelian, the Theorem 1 says that the ordinary WZW models enjoy two anomalous chiral symmetries respectively given  by the (twisted) left and ordinary  right multiplications by elements of the loop group $LK$. If $B$ is non-Abelian, the deformed WZW models still have
  two anomalous chiral Poisson-Lie symmetries.  Theorem 2 says that the left and right moment maps $\Lambda_{L}, \Lambda_R$
  can be combined into the  non-anomalous moment maps $B_{L},B_{R}$.  For  $B$ Abelian, this new moment maps are equal to each other and they generate  the adjoint action of $G$ on the target
  space of the $\si$- model.  This adjoint action is non-anomalous and serves as the base of the standard vector gauging of the WZW model leading to the gauged  $K/S$  WZW model described above. However, if $B$ is non-Abelian, the moment maps $B_L$ and $B_R$ do not coincide and we have two different non-anomalous
  quasi-adjoint actions of Theorem 2  which can be consistently gauged.  
  Finally, the Theorem 3 explains under which conditions the chiral subsymmetries may become  non-anomalous and can be consistently gauged.  As an illustration, we devote an entire Section 4 to a very explicite construction  of a  particular new deformation  of the ordinary WZW model  (which
we call the $u$-deformation) and  work out in detail its deformed vector gauging. 
 
  \medskip
  
  \noindent The paper is organized as follows: In  Section 2,  we present the discussion of the concept of the Poisson-Lie symmetry, we explain  motivations for the Definition 1 and we prove the Theorem 1. Then in Section 3.1 and 3.2, we respectively prove the Theorems 2
  and 3 and, in Section 3.3, we expose the theory of the non-decomposable doubles. In the section 4, we  construct   the $u$-deformed WZW  model  and perform its Poisson-Lie gauging. We finish with short conclusions and an outlook.

\section{Twisted Heisenberg double}

The presentation of this Section extends that of   \cite{K2}. In particular,  we give full
proofs of the statements listed in \cite{K2}, and, moreover, we are more
general concerning the properties of the twist $\k$ of a double $D$.   
\subsection{ Lie groups in a dual language}
Let $B$ be a Lie group and $Fun(B)$ the algebra of   functions on it.  It is well known that the group structure
on $B$ gives rise to a so called coproduct $\Delta:Fun(B)\to  Fun(B)\otimes Fun(B)$, the antipode
$S:Fun(B)\to Fun(B)$ and the counit $\e:Fun(B)\to \br$ given, respectively,  by the formulae
$$\Delta x(b_1,b_2)=x'(b_1)x''(b_2)=x(b_1b_2), \quad S(x)(b)=x(b^{-1}), \quad  \e(x)=x(e_B).$$
Here $x\in Fun(B)$, $b,b_1,b_2\in B$,  $e_B$ is the unit element of $B$ and we use the Sweedler notation for the coproduct: $$\Delta x= \sum_\al x'_\al\otimes x''_\al\equiv x'\otimes x''.$$ 
The Lie algebra $\B$ of $B$ is defined as the set of $\e$-derivations of $Fun(B)$, i.e.
$$\B=\{\delta: Fun(B)\to\br, \delta(xy)=\e(x)\delta(y)+\e(y)\de(x)\}.$$
The Lie bracket on $\B$ is defined as follows:
$$[\de_1,\de_2](x)=\de_1(x')\de_2(x'')-\de_1(x'')\de_2(x').$$
This definition of the Lie algebra $\B$ is of course equivalent to a more standard one presenting $\B$ 
as the set of right-invariant vector fields. In order to connect two definitions,  consider a map $\phi^B: Fun(B)\to \Omega^1(B)$ (the map $\phi^B$  thus goes from functions into 1-forms on $B$) defined by 
$$\phi^B(x)= dx'S(x'').$$
Note that the 1-form $\phi^B(x)$ is automatically right-invariant  therefore the canonical pairing of  a right-invariant vector field $v$ with $\phi^B(x)$ defines a map $\de_v: Fun(B)\to \br$:
$$\de_v(x)=<v,\phi^B(x)>. \eqno(0)$$
 The map $\de_v$ is indeed the $\e$-derivation due to the following property of the map $\phi^B$:
 $$\phi^B(xy)=\e(x)\phi^B(y) +\e(y)\phi^B(x).$$
 On the other hand, every $\e$-derivation $\de$ defines a right-invariant vector field $\nabla^L_\de$ which
 acts on $x\in Fun(B)$ as follows:
 $$\nabla^L_\de x=\de( x' )x''.$$
 \noindent    Consider now a Poisson-Lie group $B$, i.e. a Lie group equipped with a Poisson bracket
 $\{.,.\}_B$
 satisfying
 $$\Delta\{x,y\}_B=\{x',y'\}_B\otimes x''y''+x'y'\otimes \{x'',y''\}_B, \quad x,y\in Fun(B).\eqno(1)$$
It is not difficult to prove that the property (1) implies
$$S(\{x,y\}_B)=-\{S(x),S(y)\}_B, \quad  x,y\in Fun(B)\eqno(2a)$$
$$\e(\{x,y\}_B)=0, \quad x,y\in Fun(B).\eqno(2b)$$
Denote by $\B^*$  the linear dual  of the Lie algebra $\B=Lie(B)$. The Poisson-Lie bracket $\{.,.\}_B$ induces a natural Lie algebra structure $[.,.]^*$ on $\B^*$.   Let us explain this fact in more detail: 
First of all recall that $\B^*$ can be identified with the space of right-invariant $1$-forms on the group
manifold $B$ and we have the  natural (surjective)  map   $\phi^B: Fun(B)\to \B^*$   defined by 
$$\phi^B(y)= dy'S(y''), \quad y\in Fun(B).$$
Note that the $1$-form $\phi^B(y)$ is   right-invariant  therefore it is indeed in $\B^*$.  Let 
$U,V\in \B^*$ and $x,y\in Fun(B)$ such that $U=\phi^B(x)$ and $V=\phi^B(y)$. Then we define
$$[U,V]^*=\phi^B(\{x,y\}_B).\eqno(2c)$$
It is the Poisson-Lie property (1) of  $\{.,.\}_B$ which ensures the independence of $[U,V]^*$ on the
choice of the representatives $x,y$. In what follows, the Lie algebra $(\B^*, [.,.]^*)$ will be denoted 
by the symbol $\G$  and $G$ will be a (connected simply connected) Lie group such that $\G=Lie(G)$.
We note that $G$ is often referred to as the dual group of $B$. It can be itself equipped with a 
Poisson-Lie bracket $\{.,.\}_G$ inducing on $\G^*\equiv \B$ the correct Lie algebra structure $Lie(B)$.

\subsection{Poisson-Lie symmetry}
 
 The  concept of the Poisson-Lie symmetry of a symplectic manifold $M$ was introduced by
 Semenov-Tian-Shansky \cite{ST1} . Traditionally, it  concerns the  action of a Poisson-Lie group
 $G$ on $M$ such that the smooth map $G\times M\to M$ is Poisson.  Certain
 Poisson-Lie symmetries have moment maps $\mu:M\to B$, where $B$ is the  dual Poisson-Lie group. 
 Let $\Pi_M$ be the Poisson bivector corresponding to the symplectic structure on $M$,  let
 $\rho_B$ be the right-invariant Maurer-Cartan form on $B$ and let $<.,.>$ denote the canonical
 pairing between $Lie(B)$ and $Lie(G)$. Then the moment map $\mu$ is characterized by the property that the vector field  $\Pi_M(.,\mu^*<\rho_B,U>)\in Vect(M)$ generates the infinitesimal action of the element $U\in Lie(G)$ on $M$.  We have the following lemma:
 
 \medskip
 
 \noindent\underline{Lemma 1} : Let the action $G\times M\to M$ be the Poisson-Lie symmetry
 with the moment map $\mu:M\to B$ and let $w_\mu:Fun(B)\to Vect(M)$ be a map defined as
 $$w_\mu(y)=\Pi_M(.,\mu^*\phi^B(y)).$$
 Then $w_\mu$ is anti-homomorphism of the Lie algebras $Fun(B)$ and $Vect(M)$.
 
\medskip

\noindent\underline{Proof}:  Let $x,y$ be in $Fun(B)$. We know that  the right-invariant
$1$-forms $\phi^B(x)$ and $\phi^B(y)$ can be seen as the elements of $Lie(G)$, denote them
as $U$ and $V$, respectively. Then the statement of the Lemma follows from  Eq. (2c) and
from  the property of the moment map stated above.

 \rightline{\#}
 
 \noindent  In this paper, we shall  advocate a different approach to Poisson-Lie symmetry
 and we take the statement of the Lemma 1 as a definition. Thus we propose
 
 \medskip
 
  \noindent \underline{Definition 1}:  Let $M$ be a  symplectic  manifold whose algebra of smooth functions $Fun(M)$ is 
equipped with a Poisson bracket $\{.,.\}$. Let $B$ be  a Poisson-Lie group and let $\mu: M\to B$ be a smooth  map.
 To every function $y\in Fun(B)$ we can associate a vector field $w_\mu(y)\in Vect(M)$ as follows:   
  $$w_\mu(y)f=\{f,\mu^*(y')\} \mu^*(S(y'')), \quad y\in Fun(B), f\in Fun(M).\eqno(3)$$ 
We say that   $\mu$ realizes the Poisson-Lie symmetry of   $M$ if  the map 
$w_\mu$ is an anti-homomorphism of the Lie algebras  $Fun(B)$ and $Vect(M)$.  If, moreover, the map $\mu$
is Poisson, we say that the symmetry is equivariant or non-anomalous. 
 
 \rightline{\#}

  \noindent\underline{Explanations}:    If $\mu$ realizes the Poisson-Lie symmetry of $M$, the
  opposite Lie algebra of the image $Im(w_\mu)$ of the map $w_\mu$ is a
   Lie algebra that will be denoted as $\G$. If the action of the Lie algebra $\G$ on $M$ can be lifted
   to the action of a connected Lie group $G$
  (such that $Lie(G)=\G$)   we speak about global Poisson-Lie symmetry. $G$ will be then  referred to as the symmetry group of $(M,\mu)$ and $B$ as the
  cosymmetry group. Note that $G$ acts on $M$ and $B$ underlies the way how this action is expressed via the Poisson brackets.  If   there is distinguished (evolution)
  vector field $v\in Vect(M)$ leaving invariant $Im(\mu^*)$, we say that the dynamical system
  $(M,\{.,.\},v)$ is $(G,B)$-Poisson-Lie symmetric (cf. \cite{K2}). We also note that $y\in Fun(B)$
  can be interpreted as a non-Abelian (or Poisson-Lie) Hamiltonian of the vector field $w_\mu(y)$. The fact that
  $w_\mu$ is anti-homomorphism just implies a nice formula $[w_\mu(x),w_\mu(y)]=-w_\mu(\{x,y\}_B)$. If the group
  $B$ is Abelian  then  $\Delta(x)=1\ot x+x\ot 1$ and  (3) is nothing but the standard Hamiltonian formula $w_\mu(y)f=\{f,\mu^*(y)\}$. Thus the Poisson-Lie symmetry becomes the standard Hamiltonian symmetry 
  if the cosymmetry group $B$ is Abelian.
 \vskip1pc
 
\noindent  Let us note also that  the Definition 1 can be reformulated  
 by using the Maurer-Cartan form $\rho_B$ and thus avoiding to refer to the coproduct on $Fun(B)$ (this essentially amounts to replace $dy'S(y'')$ by $<\rho_B,V>$). There are two reasons that we choose the 
 formulation that uses the coproduct and the antipode.  First one  
  is not directly related to this paper, but is important in general in
perspective of quantization. Indeed, for the definition of the Hopf symmetry
the notions of coproduct and antipode are indispensable already at the level of
basic definition and the close relationship between the Poisson-Lie and Hopf
symmetry thus becomes more transparent.The second reason is more practical.
  In
fact, the notation using the coproduct and the antipode is technically more convenient in
elaborating and formulating proofs of the theorems presented in the paper.

\vskip1pc
 \noindent\underline{Remark}: Our definition of the Poisson-Lie symmetry
  and the  traditional one are close cousins but they are not quite identical.  For example, a traditional symmetry
  must admit a moment map in order to be the symmetry in the new sense and the newly defined symmetry must be global in order to be traditional.   The main reason why we shall use the new  definition  
  is its usefulness for treatment of anomalies which  cause obstructions for gauging the Poisson-Lie
  symmetries. The traditional definition does not see the difference between anomalous and non-anomalous cases while the new definition gives the very simply criterion to  distinguish them.  
  In what follows, we shall work exclusively with the new definition and we hope to convince the reader
  about its naturaleness and usefulness.

 \medskip

  \noindent\underline{Lemma 2}:  Every Poisson map  $\mu:M\to B$ realizes the Poisson-Lie symmetry of $M$.
 
 \medskip
 
  \noindent \underline{Proof}:  
  First remind that the map $\mu:M\to B$ is a  Poisson morphism iff  the dual map $\mu^*:Fun(B)\to Fun(M)$
  satisfies
  $$\{\mu^*(x),\mu^*(y)\}=\mu^*(\{x,y\}_B), \quad x,y\in Fun(B).\eqno(4)$$
  Now we take $x,y\in Fun(B)$ and  calculate 
  $$[w_\mu(y),w_\mu(x)]f =$$
  $$\{\{f,\mu^*(x')\}\mu^*(S(x''))\},\mu^*(y')\}\mu^*(S(y''))-\{\{f,\mu^*(y')\}\mu^*(S(y''))\},\mu^*(x')\}\mu^*(S(x''))=$$
  $$=\{f,\{\mu^*(x'),\mu^*(y')\}\}S(\mu^*(x''))S(\mu^*(y''))-\{f,\mu^*(x'y')\}\{\mu^*(S(x'')),\mu^*(S(y''))\}=$$
  $$=\{f,\mu^*(\{x,y\}_B')\}\mu^*(S(\{x,y\}''_B))=w_\mu(\{x,y\}_B)f$$
  Going from the second to the  third line we have used the Jacobi identity and the fact that 
  $x'S(x'')$ is a number (the counit of $x$). We have passed from the third to the fourth line
  by using (1),(2ab) and (4).  
  
  \rightline{\#}
   \vskip3pt

  \subsection{Anomalous realizations}
  The Poisson-Lie symmetry can be realized also by a map   
  $\mu:M\to B$ which is not the Poisson morphism. If this happens we speak about the  anomalous
  Poisson-Lie symmetry and we call $\mu$ the anomalous moment map.  Anomalous moment maps
naturally arise by twisting the Heisenberg doubles. The detailed exposition of this 
fact will be our following subject. 
  
\vskip1pc

\noindent \underline{Definition 2}:  Let $D$ be an even-dimensional Lie group equipped with a maximally Lorentzian
bi-invariant metric. If  
$Lie(D)=Lie(G)\stackrel{.}{+}Lie(B)$, where 
$G$ and $B$ are maximally isotropic  subgroups, $D$ is called the Drinfeld double of $G$ or the Drinfeld double of $B$.  Let $\k$ be a  metric preserving 
automorphism of $D$   and suppose that there are  respective basis  $T^i$ and $t_i$ 
($i=1,...,n$)  of $\G=Lie(G)$ and $\B=Lie(B)$ such that 
 $$(T^i,t_j)_\D=\delta^i_j.\eqno(5)$$
Then the (basis independent) expression 
$$\{f_1,f_2\}_D\equiv \nabla^R_{T^i}f_1 \nabla^R_{t_i}f_2 -
\nabla^L_{\k(t_i)}f_1\nabla^L_{\k(T^i)}f_2, \quad f_1,f_2\in Fun(D) \eqno(6) $$
is a  Poisson bracket and the Poisson manifold $(D,\{.,.\}_D)$ is called the twisted Heisenberg double.

 \vskip1pc

\noindent \underline{Theorem 1}:
\noindent Let $D$ be a  twisted Heisenberg double which is also decomposable, i.e. such that two global unambiguous decompositions hold: $D=\k(B)G$ 
and $D=\k(G)B$.  Consider  (smooth)  maps $\L_{L}, \L_R: D\to B$, $\Xi_R,\Xi_L:D\to G$   respectively induced  by these two decompositions. Then  it holds:

\medskip

\noindent a)  The Poisson manifold $(D,\{.,.\}_D)$ is   symplectic. 
 
  \noindent b) Both 
maps $\L_L$ and $\L_R$ realize the global (anomalous)  Poisson-Lie symmetries of the symplectic manifold 
$(D,\{.,.\}_D)$.  The corresponding symmetry group is $G$ acting  as 
$$h\tr K=\k(h)K , \quad h\in G,\quad  K\in D, \eqno(7a)$$
or, respectively, as
 $$h\tr K=Kh^{-1} , \quad h\in G,\quad  K\in D.\eqno(7b)$$
\vskip1pc

\noindent \underline{Explanations}:   The symbol $\stackrel{.}{+}$ stands for the direct sum of vector spaces only and
not of Lie algebras. Bi-invariant  means both left-  and right-invariant. The  non-degenerated  bi-invariant metric on $D$
obviously  induces an $Ad$-invariant non-degenerated bilinear form  $(.,.)_{\D}$ on $\D=Lie(D)$.
 An  isotropic submanifold  of $D$ is such that the induced metric on it vanishes. Maximally isotropic
 means  that it is not contained in any bigger isotropic submanifold. The  vector fields 
 $\nabla^{L,R}_T$ are defined as  
 $$\nabla^L_T f(K)\equiv \delta_T(f')f''(K)=\biggl({d\over ds}\biggr)_{s=0}f(e^{sT}K), $$
 $$ \nabla^R_T f(K)\equiv \delta_T(f'')f'(K)=\biggl({d\over ds}\biggr)_{s=0}f(Ke^{sT}),$$
where  $f\in Fun(D)$, $K\in D,T\in Lie(D).$
\noindent   Global unambiguous decomposition $D=\k(B)G$ means that for every 
element $K\in D$ it exists a unique $g=\Xi_R(K)\in G$ and a unique $b=\L_L(K)\in B$ such that
 $K=\k(b) g^{-1}$. Similarly for $D=\k(G)B$:
it exists a unique $\ti g=\Xi_L(K)\in G$ and a unique $\ti b=\L_R(K)\in B$ 
such that  $K=\k(\ti g)\ti b^{-1}$. 
The fact that the formula (6) defines the Poisson bracket  was proved by Semenov-Tian-Shansky in \cite{ST2} and, for completeness, we shall   outline here his argument: 

 \medskip
 
 \noindent  Consider a (basis independent) element $c\in \D\ot \D$ given by
 $$c=T^i\ot t_i+t_i\ot T^i.$$
 It is easy to see that the $Ad$-invariance and $\k$-invariance of the bilinear form $(.,.)_\D$ implies the $Ad$-invariance and $\k$-invariance of $c$.  Thus
 the bracket (6) can be rewritten as 
$$\{f_1,f_2\}_D=\jp \nabla^R_{T^i}f_1 \nabla^R_{t_i}f_2 -    \jp \nabla^R_{t_i}f_1 \nabla^R_{T^i}f_2          +\jp \nabla^L_{\k(T^i)}f_1 \nabla^L_{\k(t_i)}f_2 -    \jp \nabla^L_{\k(t_i)}f_1 \nabla^L_{\k(T^i)}f_2.   $$
Note that in this bracket appear two    elements of $\D\w \D$ given by
$$r_\D = \jp  T^i\ot t_i-\jp t_i\ot T^i, \quad r_\D^\k =\jp \k(T^i)\ot \k(t_i)-\jp \k(t_i)\ot \k(T^i).$$
It can be shown by direct calculation that the algebraic Schouten brackets $[r_\D,r_\D]_S$
(cf. \cite {K}, Eqs. (4.36-39) ) gives
an invariant element of $\w^3\D$ and, moreover, $[r^\k_\D,r^\k_\D]_S$= $[r_\D,r_\D]_S$. Those facts
imply that the Semenov-Tian-Shansky bracket (6) satisfies the Jacobi identity.

 \vskip1pc
 
 \noindent  Let us finish the Explanations by saying that the list of decomposable doubles 
 is not very long. The typical examples are the cotangent bundle $T^*G$ of  any Lie group $G$, the complexification $G^\bc$ of a compact (loop)
 group $G$ and certain Drinfeld twists of two first items.   Nevertheless,  
the independent theorem dealing with decomposable doubles   is useful for two
reasons. First of them is the range of applicability: many resoluble quantum
theories have compact (quantum) group symmetry and in this or other way are
based on the short list of decomposable doubles. The other reason is that the
notion of the Poisson-Lie symmetry is traditionally globally defined and the
decomposable  doubles lead to global Poisson-Lie symmetry.  Let us stress, however,
that the local Poisson-Lie symmetries
must be considered equally seriously (for instance the conformal symmetry in
  field theory is only local but physically relevant). This is the reason that
we devote   the section 3.3 to non-decomposable doubles where the number of
examples is very big.

\vskip1pc

\noindent \underline{Proof of Theorem 1}:
 
 \medskip
 
\noindent{a)}: Consider a point $K\in D$ and four linear subspaces
 of the tangent space $T_KD$ defined as $S_L=L_{K*}\G$, $S_R=R_{K*}\k(\G)$,
 $\tilde S_L=L_{K*}\B$ and  $\tilde S_R=R_{K*}\k(\B)$. (The symbols $L_{K*}$ and $R_{K*}$
stand for left and right transport on the group $D$, respectively). The existence of the global 
decompositions $D=\k(B)G$ and $D=\k(G)B$ means that at every $K\in D$ the tangent space
$T_KD$ can be decomposed as $T_KD=S_L+\ti S_R$ and $T_KD=\ti S_L +S_R$, respectively.
This fact makes possible to introduce a projector $\Pi_{L\ti R}$ on $\ti S_R$ with a
kernel $S_L$ and a projector $\Pi_{\ti L R}$ on $S_R$ with a
kernel $\ti S_L$.  At every point $K\in D$ we can therefore define a following
 2-form $\om$ 
$$\om (t,u)=(t,(\Pi_{\ti L R}-\Pi_{L\ti R})u)_{\D}, \eqno(8)$$
where $t,u$ are arbitrary   vectors in $T_KD$ and    $(.,.)_{\D}$ is the bi-invariant
metric at the point $K$ (it is related by the left or right transport of the Ad-invariant
bilinear form $(.,.)_{\D}$ defined at the unit element $E\in D$). Let us show that $\om$ 
is the symplectic form corresponding to the Poisson structure $\{ .,.\}_D$. First
of all we remark that the Poisson bivector (=contravariant antisymmetric tensor) 
corresponding to the Poisson bracket $\{ .,.\}_D$ reads
$$\al=L_{K*}(T^i\ot t_i)- R_{K*}(\k(t_i)\ot \k(T^i)).\eqno(9)$$
Introduce two more projectors $\Pi_{R\ti R},\Pi_{\ti LL}$, where the first
subscript stands for the kernel and the second for the image. Then we conclude
$$\al(.,\om(.,u))=$$ $$=L_{K*}T^i(L_{K*}t_i,(\Pi_{\ti LR}-\Pi_{L\ti R})u)_{\D}-
R_{K*}\k(t_i)(R_{K*}\k(T^i),(\Pi_{\ti LR}-\Pi_{L\ti R})u)_{\D}=$$
$$=(\Pi_{\ti LL}-\Pi_{R\ti R})(\Pi_{\ti LR}-\Pi_{L\ti R})u=u.\eqno(10)$$ 
 
 \vskip3pt
\vskip1pc
\noindent\underline{Proof of b) and c)}
 Consider a  bracket $\{.,.\}_B$ on the cosymmetry
group $B$ given by
$$\{x,y\}_B(b)=-(T^i,Ad_b T^k)_{\D}(\nabla^L_{t_i}x)(b)
 (\nabla^R_{t_k}y)(b),
 \quad b\in B , \quad x,y\in Fun(B).\eqno(11)$$
It was shown in Proposition 4.5. of \cite{K} that $\{.,.\}_B$ is the Poisson-Lie
bracket on $B$. We shall prove that
$$\{\L_L^*(x),\L_L^*(y)\}_D=\L_L^*\biggl(\{x,y\}_B - M_\k^{ij}\nabla^R_{t_i}x \nabla^R_{t_j}y\biggr),\quad x,y \in Fun(B),\eqno(12a)$$
$$\{\L_R^*(x),\L_R^*(y)\}_D=\L_R^*\biggl(\{x,y\}_B -  M^{ij}_{\k^{-1}}\nabla^R_{t_i}x \nabla^R_{t_j}y \biggr), \quad x,y \in Fun(B),\eqno(12b)$$
where  the constant antisymmetric matrix $M_\k^{ij}$  is given by 
$$M_{\k}=Q_\k P_\k^{-1}, \quad  (P_\k )_i^{\ j}=(\k(t_i), T^j)_{\D},\quad Q_\k^{ij}=(\k(T^i),T^j)_{\D}.\eqno(13)$$
We note that the non-degeneracy of $(.,.)_\D$ and also the  global decomposabilities  $D=\k(B)G=\k(G)B$ guarantee that both matrices $P_\k$  and
$P_{\k^{-1}}$ are  invertible. 

\vskip1pc
\noindent  In order to calculate the bracket $\{\L_L^*(x),\L_L^*(y)\}_D$, we use the defining
formula (6). We first  realize that
$$ \nabla^R_{T^i}\L_L^*(x)=\biggl({d\over ds}\biggr)_{s=0}x(\L_L(Ke^{sT^i}))=0\eqno(14)$$
and then we write
$$\{\L_L^*(x),\L_L^*(y)\}_D=-\nabla^L_{\k(t_i)}\L_L^*(x)\nabla^L_{\k(T^i)}\L_L^*(y)=$$
$$=-\biggl({d\over ds_1}\biggr)_{s_1=0}x(\L_L(e^{s_1\k(t_i)}K)) 
\biggl({d\over ds_2}\biggr)_{s_2=0}y(\L_L(e^{s_2\k(T^i)}K))=$$
$$=-\L_L^*(\ ^B\nabla^L_{t_i}x)\biggl({d\over ds_2}\biggr)_{s_2=0}y(\L_L(e^{s_2\k(T^i)}\k(\L_L(K))))=$$
$$=-\L_L^*(\ ^B\nabla^L_{t_i}x)\biggl({d\over ds}\biggr)_{s=0}y(\L_L(\k[\L_L(K)\exp{(s\L_L^{-1}(K)T^i\L_L(K))}])).\eqno(15)$$
We note that
$$\L_L^{-1}(K)T^i\L_L(K)=(\L_L^{-1}(K)T^i\L_L(K),t_k)_{\D}T^k +(\L_L^{-1}(K)T^i\L_L(K), T^k)_{\D}t_k.$$
This identity permits to  rewrite the r.h.s. of (15) as the sum of two terms
$$\{\L_L^*(x),\L_L^*(y)\}_D=V_1+V_2,$$
where
$$V_1= - (\L_L^{-1}(K)T^i\L_L(K), T^k)_{\D}\L_L^*(\ ^B\nabla^L_{t_i}x) \L_L^*(\ ^B\nabla^R_{t_k}y)
= \L_L^*(\{x,y\}_B)$$
and
$$V_2= 
- (\L_L^{-1}(K)T^i\L_L(K), t_k)_{\D}\L_L^*(\ ^B\nabla^L_{t_i}x)  \biggl({d\over ds}\biggr)_{s=0}y(\L_L(\k[\L_L(K)\exp{(sT^k)}]))=$$
$$-\L_L^*(\ ^B\nabla^R_{t_k}x)\biggl({d\over ds}\biggr)_{s=0}y(\L_L(\k[\L_L(K)\exp{(s\tau^k)}]))=
-\L_L^*(\ ^B\nabla^R_{t_k}x)\L_L^*(\ ^B\nabla^R_{\tau^k}y).$$
The element $\tau^k\in \B$ is defined by the $D=\k(B)G$ decomposition 
$$\k(T^k)=\k(\tau^k)+ c^k, \quad c^k\in \G.$$
From this it is easy to find that
$$\tau^k= M_\k^{kl}t_l,$$
where the matrix $M_\k$ was introduced in (13). Putting all together, we arrive at
$$\{\L_L^*(x),\L_L^*(y)\}_D=\L_L^*\biggl(\{x,y\}_B - M_\k^{ij}\ ^B\nabla^R_{t_i}x\ ^B \nabla^R_{t_j}y\biggr),$$
which is nothing but (12a).  The identity (12b) can be proved in a similar way.  We note also that
our notation has  distinguished   the invariant derivatives on $Fun(D)$ and on $Fun(B)$
(the derivatives on $Fun(B)$ where denoted as $\ ^B\nabla^{R,L}$). We shall not make this distinction
in what follows and we let the reader to understand from the context on which space $\nabla^{R,L}$ act.
\vskip1pc

\noindent  In case where the  twisting automorphism is trivial (i.e. $\k$ is identity), the
anomaly matrices $M_\k$, $M_{\k^{-1}}$ vanish and  $\L_{L,R}: D\to B$ are the Poisson maps.
From Lemma 2 it then follows that $\L_{L,R}: D\to B$ realize the Poisson-Lie symmetries of $D$.   Let us show now that in the case of non-trivial twisting 
the maps $\L_{L,R}:D\to B$ also realize the Poisson-Lie symmetries  although they are not Poisson morphisms. For this, we  first remind the definition (3) of the map $w_{\L_L}:Fun(B)\to Vect(D)$:
$$w_{\L_L}(x)f =\{f,\L_L^*(x')\}_D\L_L^*(S(x'')),\quad x\in Fun(B),\ f\in Fun(D).$$ 
We calculate
$$  (w_{\L_L}(y)w_{\L_L}(x)-w_{\L_L}(x)w_{\L_L}(y))f=$$
$$= \lpb\lpb f,\L_L^*(x')\}_D\L_L^*(S(x'')),\L_L^*(y')\}_D \L_L^*(S(y'')) -(x\leftrightarrow y)=$$
$$=\{\{f,\L_L^*(x')\}_D\L_L^*(y')\}_D\L_L^*(S(x''y''))+\{f,\L_L^*(x')\}_D\{\L_L^*(S(x'')),\L_L^*(y')\}_D\L_L^*(S(y''))$$
$$-(x\leftrightarrow y)=$$
$$=\{f,\{\L_L^*(x'),\L_L^*(y')\}_D\}_D\L_L^*(S(x''y'')) -\{f,\L_L^*(x'y')\}_D\{\L_L^*(S(x''),\L_L^*(S(y''))\}_D.$$
Now we use the formula (12a) and the Poisson-Lie property (1) of the bracket $\{.,.\}_B$ to obtain
$$[w_{\L_L}(y),w_{\L_L}(x)]f=$$
$$= \{f, \L_L^*(\{x',y'\}_B)\}_D\L_L^*(S(x''y''))-\{f,\L_L^*(x'y')\}_D\L_L^*(\{S(x''),S(y'')\}_B)+$$
$$-M^{ij}_\k\biggl( \{f, \L_L^*(\nabla^R_{t_i}x' \nabla^R_{t_j}y')\}_D\L_L^*(S(x''y'')) -\{f,\L_L^*(x'y')\}_D\L_L^*(\nabla^R_{t_i}S(x'') \nabla^R_{t_j}S(y''))\biggr)$$
The last line of this expression vanishes due to following identities
$$(\nabla^R_{t_l}y')S(y'')+y'\nabla^R_{t_l}S(y'')=\nabla^R_{t_l}(y'S(y''))=0,$$
$$(\nabla^R_{t_l}\nabla^L_{t_i}x')S(x'')+\nabla^L_{t_i}x'\nabla^R_{t_l}S(x'')=
\nabla^R_{t_l}(\nabla^L_{t_i}x'S(x''))=0$$
and (using (6))
$$\{f,\L_L^*(\nabla^R_{t_l}x')\}_D\L_L^*(S(x''))+\{f,\L_L^*(x')\}_D\L_L^*(\nabla^R_{t_l}S(x''))=$$
$$=\nabla^L_{\k(T^i)}f\L_L^*((\nabla^R_{t_l}\nabla^L_{t_i}x')S(x'')+\nabla^L_{t_i}x'\nabla^R_{t_l}S(x''))=0.$$
Now we use the Poisson-Lie properties (1),(2) to arrive at  
$$[w_{\L_L}(y),w_{\L_L}(x)]f=$$
$$= \{f, \L_L^*(\{x',y'\}_B)\}_D\L_L^*(S(x''y''))+
\{f,\L_L^*(x'y')\}_D\L_L^*(S(\{x'',y''\}_B))=$$ $$=w_{\L_L}(\{x,y\}_B)f.$$
According to the Definition 1, the map $\L_L$ thus realizes the Poisson-Lie symmetry of $D$.
\vskip1pc

\noindent Much in the same way, we obtain also
$$[ w_{\L_R}(y), w_{\L_R}(x)]f= w_{\L_R}(\{x,y\}_B)f,$$
where 
$$  w_{\L_R}(x)f=\{f,\L_R^*(x')\}_D \L_R^*(S(x'')), \quad x\in Fun(B), f\in Fun(D).$$ 
Having established that both maps $w_{\L_L}, w_{\L_R}: Fun (B)\to Vect(D)$ are Lie algebra homomorphisms
(i.e. that both $\L_L,\L_R:D\to B$ realize Poisson-Lie symmetries), it remains to find
 what are the
corresponding symmetry groups.  We use (6) and (0)
  to obtain
$$ w_{\L_L}(y)f=\{f,\L_L^*(y')\} \L_L^*(S(y''))=
\nabla^L_{\k(T^i)}f\L_L^*(( \nabla^L_{t_i}y')S(y''))=
 \delta_{t_i}(y)\nabla^L_{\k(T^i)}f.\eqno(16a)$$
We remind that $\de_{t_i} $ is the $\e$-derivative (cf. Sec. 2.1) hence  $\de_{t_i}(y)$ is a real number for every $i$. It therefore follows that $Im(w_{\L_L})=\k(\G)$ and we have proved (7a). Similarly, we  obtain
$$w_{\L_R}(y)f= -\de_{t_i}(y)\nabla^R_{T^i}f,\eqno(16b)$$
which proves (7b). 
 
\rightline{\#}
 \section{Non-anomalous  moment maps}
 
 Non-anomalous Poisson-Lie symmetries   play very important role in the symplectic
 geometry since they permit to perform the so called symplectic reduction (or "gauging"
 in the  terminology of physicists).    However, given a decomposable  twisted Heisenberg double $(D,\k)$, the basic moment maps
 $\L_L,\L_R$ are generically anomalous and cannot be gauged.  Indeed, the anomaly matrices $M_\k^{ij}, M^{ij}_{\k^{-1}}$    vanish only in the case where the twisting  automorphism $\k$ preserves the \underline{symmetry} group $G$ (cf. (13)).  In this section, we shall look for other moment
 maps (distinct from $\L_L,\L_R$) which would allow  us to gauge
 $(D,\k)$.  It turns out, that the existence of the 
 non-anomalous Poisson-Lie moment maps associated to the twisted Heisenberg double  heavily depend on the details
 of the structure of $(D,\k)$.  In the three following subsections, we  shall discuss three interesting cases, where the non-anomalous moment maps can be constructed. We shall keep the  exposition of the two first cases (a quasi-adjoint action and a  proper subsymmetry) in an abstract  level since the concrete examples will be discussed in the subsequent Section 4. However, we shall illustrate
 the third case (an improper
 subsymmetry) already in this Section 3, since later we shall not consider it anymore. 
 
 \subsection{Quasi-adjoint action}
 
 In this subsection, we shall consider the decomposable twisted Heisenberg doubles for
 which the twisting  automorphism $\k$ preserves the \underline{cosymmetry} group $B$.
 We have the following theorem:
 
 \medskip
 
 \noindent \underline{Theorem 2}:
\noindent Let $D$ be a  decomposable twisted Heisenberg double such that  the twisting automorphism $\k$ 
preserves the subgroup $B$.   Consider the anomalous moment maps $\L_L,\L_R$ and construct two new maps $B_L:D\to B$
 and $B_R:D\to B$ as follows
 $$B_L(K)=\k(\L_L(K))\L_R(K),\quad B_R(K)=\k^{-1}(\L_R(K))\L_L(K), \quad K\in D.$$
 Then it holds:  Both  maps $B_L$ and $B_R$ are Poisson and they  realize global non-anomalous Poisson-Lie 
symmetries of $(D,\{.,.\}_D)$. The corresponding symmetry group is $G$ acting  as 
$$h\tr K = \k(h)K\Xi_R(\k[h\L_L(K)]), \quad h\in G,\ K\in D,$$
or, respectively, as
$$h\tr K =\k[\Xi_L^{-1}(\L_R^{-1}(K)h^{-1})]Kh^{-1}.\quad h\in G, \ K\in D.$$
\vskip1pc
 
\noindent\underline{Proof}:  Consider two functions $x,y\in Fun(B)$.
 We know already that it holds
$$\{\L_L^*(x),\L_L^*(y)\}_D=\L_L^*\biggl(\{x,y\}_B-M_\k^{ij}\nabla^R_{t_i}x \nabla^R_{t_j}y\biggr),\quad x,y \in Fun(B),\eqno(12a)$$
$$\{\L_R^*(x),\L_R^*(y)\}_D=\L_R^*\biggl(\{x,y\}_B- M^{ij}_{\k^{-1}}\nabla^R_{t_i}x \nabla^R_{t_j}y \biggr), \quad x,y \in Fun(B), \eqno(12b)$$
where the Poisson-Lie bracket $\{.,.\}_B$ and matrices $M_\k$, $M_{\k^{-1}}$
 were defined in (11)
and in (13), respectively. Introduce maps
$\Gamma_{L}:D\to B$, $\Gamma_{R}:D\to B$  by
$$\Gamma_{L}(K)=\k(\L_L(K)), \quad \Gamma_{R}(K)=\k^{-1}(\L_R(K)), \quad K\in D$$
hence $B_L=\Gamma_L\L_R$ and $B_R=\Gamma_R\L_L$. We shall now prove that
$$\{\Ga_L^*(x),\Ga_L^*(y)\}_D=\Ga_L^*\biggl(\{x,y\}_B+M_{\k^{-1}}^{ij}\nabla^L_{t_i}x
 \nabla^L_{t_j}y\biggr),\quad x,y \in Fun(B),\eqno(17a)$$
$$\{\Ga_R^*(x),\Ga_R^*(y)\}_D=\Ga_R^*\biggl(\{x,y\}_B+M^{ij}_{\k}\nabla^L_{t_i}x \nabla^L_{t_j}y \biggr), \quad x,y \in Fun(B).\eqno(17b)$$
First we remark that
$$(\nabla^R_{T^i}\L^*_L( x))(K)=
\biggl({d\over ds}\biggr)_{s=0}x\biggl((\L_L(Ke^{sT^i}))\biggr) =0,
 \quad K\in D,$$
$$(\nabla^L_{\k(T^i)}\Ga^*_R( y))(K)=
\biggl({d\over ds}\biggr)_{s=0}y\biggl(\k^{-1}(\L_R(e^{s\k(T^i)}K))\biggr) =0,
 \quad K\in D.$$
Thus, using the fundamental definition (6), we obtain
$$\{\L_L^*(x),\Ga_R^*(y)\}_D=0$$
and
$$\{\Ga_R^*(x),\Ga_R^*(y)\}_D= $$
$$= \biggl({d\over ds_1}\biggr)_{s_1=0}x\biggl(\k^{-1}(\L_R(Ke^{s_1T^i}))\biggr) \biggl({d\over ds_2}\biggr)_{s_2=0}y\biggl(\k^{-1}(\L_R(Ke^{s_2t_i}))\biggr) =$$
$$=- \biggl({d\over ds}\biggr)_{s=0}x\biggl(\k^{-1}(\L_R(Ke^{sT^i}))\biggr)\Ga_R^* \biggl(\nabla^L_{\k^{-1}(t_i)}y\biggr)=$$
$$=\Ga_R^*\biggl((b^{-1}\k^{-1}(T^i)b,T^j)_{\D}\nabla^R_{t_j}x\nabla^L_{\k^{-1}(t_i)}y\biggr)=$$
$$=\Ga_R^*\biggl(\biggl[(b^{-1} T^i b,T^j)_{\D}-(T^i,\k^{-1}(T^m))_{\D}(bT^jb^{-1},t_l)_{\D}(T^l,
\k^{-1}(t_m))_{\D}\biggr]\nabla^R_{t_j}x\nabla^L_{t_i}y\biggr) =$$
$$=\Ga_R^*\biggl(\{x,y\}_B+(T^i,\k^{-1}(T^m))_{\D}(\k^{-1}(t_m),T^j)_{\D}\nabla^L_{t_i}x\nabla^L_{t_j}y\biggr)=$$
$$=\Ga_R^*\biggl(\{x,y\}_B+M_{\k}^{ij}\nabla^L_{t_i}x \nabla^L_{t_j}y\biggr).\eqno(17b)$$
We note that $b\in B$ in this formula denotes the argument of functions in $Fun(B)$.
Similarly, we can prove that
$$\{\L_R^*(x),\Ga_L^*(y)\}_D=0$$
and 
$$\{\Ga_L^*(x),\Ga_L^*(y)\}_D=\Ga_L^*\biggl(\{x,y\}_B+M_{\k^{-1}}^{ij}\nabla^L_{t_i}x
 \nabla^L_{t_j}y\biggr),\quad x,y \in Fun(B),\eqno(17a)$$
 
\noindent Now we calculate
$$\{B^*_L(x),B^*_L(y)\}_D=\{\Ga^*_L(x')\L_R(x''),\Ga^*_L(y')\L_R(y'')\}_D=$$
$$=\{\Ga^*_L(x'),\Ga^*_L(y')\}\L_R(x'')\L_R(y'')+
\Ga^*_L(x')\Ga^*_L(y')\{\L_R(x''),\L_R(y'')\}=$$
$$\Ga_L^*\biggl(\{x',y'\}_B+M_{\k^{-1}}^{ij}\nabla^L_{t_i}x'
 \nabla^L_{t_j}y'\biggr)\L_R(x'')\L_R(y'')+$$ $$+\Ga^*_L(x')\Ga^*_L(y')
\L_R^*\biggl(\{x'',y''\}_B- M^{ij}_{\k^{-1}}
\nabla^R_{t_i}x'' \nabla^R_{t_j}y'' \biggr)=$$ 
$$=B^*_L\biggl(\{x,y\}_B+M^{ij}_{\k^{-1}}\nabla^L_{t_i}x\nabla^L_{t_j}y  
-M^{ij}_{\k^{-1}}\nabla^R_{t_i}x\nabla^R_{t_j}y\biggr),\eqno(18a)$$
Similarly, we obtain
$$\{B^*_R(x),B^*_R(y)\}_D=B^*_R\biggl(\{x,y\}_B+M^{ij}_{\k}
\nabla^L_{t_i}x\nabla^L_{t_j}y  
-M^{ij}_{\k}\nabla^R_{t_i}x\nabla^R_{t_j}y\biggr).\eqno(18b)$$
The reader may be surprised by the presence of the anomaly matrices 
$M_\k$, $M_{\k^{-1}}$ in the resulting formulas (18a) and (18b). Didn't we promise
that the moment maps $B_L,B_R$ realize non-anomalous Poisson-Lie symmetries?
Well, the point is the following: If the twisting automorphism $\k$ preserves the
cosymmetry group $B$ then there are \underline{three} natural Poisson-Lie brackets on $Fun(B)$.
The first one is evident; it is given by the formula (11) of Section 2.3:
$$\{x,y\}_B(b)=-(T^i,Ad_b T^k)_{\D}(\nabla^L_{t_i}x)(b)
 (\nabla^R_{t_k}y)(b),
 \quad b\in B , \quad x,y\in Fun(B).$$
The second and the third bracket are defined by
$$\{x,y\}_B^{\k}(b)=-(\k(T^i),Ad_b \k(T^k))_{\D}(\nabla^L_{\k (t_i)}x)(b)
 (\nabla^R_{\k(t_k)}y)(b),\eqno(19a)$$
$$\{x,y\}_B^{\k^{-1}}(b)=-(\k^{-1}(T^i),Ad_b \k^{-1}(T^k))_{\D}(\nabla^L_{\k^{-1} (t_i)}x)(b)
 (\nabla^R_{\k^{-1}(t_k)}y)(b).\eqno(19b)$$
It is easy to understand why the brackets (19a) and (19b) verify the Jacobi identity 
and the Poisson-Lie property (1). It is because they appear on the same footing as the
original bracket (11). Indeed, the double $D$ is not only the double of the pair of
groups $G$ and $B$, but it is also the double of the pair $\k(G)$ and $ \k(B)=B$  and of the
pair $\k^{-1}(G)$ and $ \k^{-1}(B)=B$. Each of the three pairs generate the respective basis
$T^i,t_i$; $\k(T^i),\k(t_i)$ and $\k^{-1}(T^i),\k^{-1}(t_i)$, all three basis sharing
 the crucial duality property (5).
\vskip1pc

\noindent The brackets (19a) and (19b) can be worked out in the basis $t_i$ instead of
$\k(t_i)$ or $\k^{-1}(t_i)$. We use obvious identities
$$\k(t^i)=(\k(t^i),T^m)_\D t_m, \quad  \k^{-1}(t^i)=(\k^{-1}(t^i),T^m)_\D t_m$$
and we find 
$$\{x,y\}_B^{\k}=\{x,y\}_B+M^{ij}_{\k^{-1}}
\nabla^L_{t_i}x\nabla^L_{t_j}y  
-M^{ij}_{\k^{-1}}\nabla^R_{t_i}x\nabla^R_{t_j}y,$$
$$\{x,y\}_B^{\k^{-1}}=\{x,y\}_B+M^{ij}_{\k}\nabla^L_{t_i}x\nabla^L_{t_j}y  
-M^{ij}_{\k}\nabla^R_{t_i}x\nabla^R_{t_j}y.$$
This permits us to rewrite (18a) and (18b) as
$$\{B^*_L(x),B^*_L(y)\}_D=B^*_L\biggl(\{x,y\}^{\k}_B\biggr),$$ 
$$\{B^*_R(x),B^*_R(y)\}_D=B^*_R\biggl(\{x,y\}^{\k^{-1}}_B\biggr).$$ 
We thus conclude that the moment maps $B_L$ and $B_R$ are indeed non-anomalous
with respect to the Poisson-Lie brackets (19a) and (19b).
 
\vskip1pc
\noindent  Every Poisson-Lie moment map $\mu$ generates the action of the Lie algebra $\G$
 and, in good cases, this $\G$-action can be lifted to the
action of the symmetry group $G$.   Let us now show that the moment maps $B_L,B_R$ are
those "good" cases yielding the global non-anomalous Poisson-Lie symmetries. The following
exposition uses some standard conventions concerning the Hopf algebra calculations (see \cite{Kas}), namely,
the repeated application of the coproduct is written as
$$(\Delta\ot Id\ot Id)(\Delta\ot Id)\Delta(x)\equiv x'\ot x''\ot x'''\ot x'''', \quad x\in Fun(B).$$
The reader has certainly noticed that this is the generalization of the Sweedler notation introduced in
Section 2.1. 
\vskip1pc
\noindent Consider first a set of functions $x^i\in Fun(B)$ which is dual to the basis $t_i$ of $\B=Lie(B)$, i.e. it holds
$$\delta_{t_j}(x^j)=\delta^i_j ,$$
where $\delta_{t_j}$ are the $\e$-derivatives.  We denote by $\k(x^i)$  the functions on $B$ of the form
$$\k(x^i)(b)=x^i(\k(b)), \quad b\in B.$$ We are going to  make explicit the  basic map $w_{B_L}: Fun(B)\to Vect(D)$ expressing the action of $Lie(G)$ on $f\in Fun(D)$ (cf. (3)).
$$w_{B_L}(\k^{-1}(x^i))f= \{f,B^*_L(\k^{-1}((x^i)'))\}B^*_L(S(\k^{-1}((x^i)'')))=$$
$$\{f, \Ga^*_L(\k^{-1}((x^i)'))\L^*_R(\k^{-1}((x^i)''))\}\Ga^*_L(S(\k^{-1}((x^i)''''))))\L^*_R(S(\k^{-1}((x^i)''')))=$$
$$=\nabla^L_{\k(T^i)}f-\delta_{t_k}(\k^{-1}((x^i)''))\Ga^*_L(\k^{-1}((x^i)'))\Ga^*_L(S(\k^{-1}((x^i)''')))\nabla^R_{T^k}f=$$
$$=\nabla^L_{\k(T^i)}f- (\Ga_L(K)t_k \Ga_L^{-1}(K),\k(T^i))_\D  \nabla^R_{T^k}f=$$
$$=\nabla^L_{\k(T^i)}f- (\L_L(K)\k^{-1}(t_k) \L_L^{-1}(K),T^i)_\D  \nabla^R_{T^k}f.$$
Similarly, we obtain 
$$w_{B_R}(\k(x^i))f= \{f,B^*_R(\k((x^i)'))\}B^*_R(S(\k((x^i)''))=$$
$$=-\nabla^R_{T^i}f +  (\L_R(K)\k(t_k) \L_R^{-1}(K),T^i)_\D  \nabla^L_{\k(T^k)}f.$$
Note that $K\in D$ stands for the argument of the functions from $Fun(D)$. 

\vskip1pc
\noindent  The Lie algebra $\G$-actions can be lifted to the group  $G$-actions. The corresponding formulae can
be written in a compact form by using the maps defined by the global decompositions $D=\k(G)B$ and $D=\k(B)G$.
On the top of the maps $\L_L,\L_R:D\to B$ we have also the maps $\Xi_L,\Xi_R:D\to G$ respectively  defined as 
$K=\k(\Xi_L(K))\L_R^{-1}(K)$ and  $K=\k(\L_L(K))\Xi_R^{-1}(K)$, $K\in D$.  The actions of $\G$ on $D$ via the
vector fields $w_{B_L}(\k^{-1}(x^i))$ and $w_{B_R}(\k(x^i))$  is then respectively lifted to the $G$-actions as follows
$$h\tr K = \k(h)K\Xi_R(\k[h\L_L(K)]), \quad h\in G,\ K\in D,\eqno(20a)$$
$$h\tr K =\k[\Xi_L^{-1}(\L_R^{-1}(K)h^{-1})]Kh^{-1}.\quad h\in G, \ K\in D.\eqno(20b)$$
It is easy to verify that, in both cases, it holds:
$$(h_1h_2)\tr K=h_1\tr (h_2\tr K).$$
In particular,   when the cosymmetry group $B$ is 
Abelian,  the $G$-actions induced by the moment maps
$B_L$ and $B_R$ coincide and give  nothing but the twisted adjoint action of $G$ on $D$ (i.e. $h\tr K = \k(h)Kh^{-1}$, $h\in G$, $K\in D$). This fact, that will be proved in Section 4, justifies our terminology "quasi-adjoint" action for the case of non-Abelian
cosymmetry groups. 

\rightline{\#}
 
 \subsection{Proper subsymmetry}
  
In the case of the standard Hamiltonian symmetry, \underline{every} subgroup $H$ of 
the symmetry group $G$ also realizes the Hamiltonian symmetry. In the general
Poisson-Lie context (anomalous or not), such   statement is generically false. 
A natural question then arises: which  subgroups of $G$  are themselves Poisson-Lie
symmetry groups? We are going to answer this question  and we  also  determine the corresponding 
moment maps.
 \vskip1pc
\noindent \underline{Theorem 3}:  Let $D$ be a  decomposable twisted Heisenberg double, $\kappa$ an automorphism
of $D$ preserving  $B$ and $N$   a normal subgroup  of $B$. Denote by  $C$  the   factor group $B/N$,  by    $\rho$   the natural homomorphism $ B\to C$ and by $P_\k:Lie(D)\to Lie(B)$ a projector on $Lie(B)$ with kernel
$\k(Lie(G))$.    Suppose that the  Hopf subalgebra  $\rho^*(Fun(C))$ of $Fun(B)$  is also a Poisson subalgebra.    
 Then it holds:   The  composed map
 $\nu_R\equiv \rho\circ \L_R$ realizes  Poisson-Lie symmetry of $D$ and the corresponding 
 symmetry group $H$ is   the subgroup of $G$.   If, moreover,  $P_\k(Lie(H))\subset Lie(N)$ then the moment map $\nu_R$ is non-anomalous.

\vskip1pc

\noindent\underline{Proof}: The Poisson-Lie bracket on $Fun(B)$ naturally 
induces the Poisson-Lie bracket on $Fun(C)$ because $\rho^*(Fun(C))$ is the
 Poisson subalgebra of $Fun(B)$. Thus 
$$\{\rho^*(u),\rho^*(v)\}_B=\rho^*(\{u,v\}_C), \quad u,v\in Fun(C).$$
Now define
$$w_{\nu_R}(u)f\equiv \{f,\nu_R^*(u')\}_D\nu_R^*(S_C(u'')), \quad u\in Fun(C), f\in Fun(D)$$ 
  and calculate 
$$w_{\nu_R}(\{u,v\}_C)= \{f,\nu_R^*(\{u,v\}_C')\}_D\nu_R^*(S_C(\{u,v\}''_C))= $$
 $$=\{f,\L_R^*(\{\rho^*(u),\rho^*(v)\}'_B)\}_D\L_R^*(S_B(\{\rho^*(u),\rho^*(v)\}''_B))=$$
 $$=w_{\L_R}(\{\rho^*(u),\rho^*(v)\}_B)= [w_{\L_R}(\rho^*(u)),w_{\L_R}(\rho^*(v))]=
[w_{\nu_R}(u), w_{\nu_R}(v)].$$
Here we have used the  obvious fact  that
 $$w_{\nu_R}(u)=w_{\L_R}(\rho^*(u)).$$
 This fact also directly implies, that $H$ is the subgroup of $G$.

 \medskip
 
 \noindent Let us see  how the Lie algebra $Lie(H)$ of $H$ is located
in the Lie algebra $Lie(D)$ of the double $D$. Choose a  vector
subspace $V\subset Lie(B)$ that is complement  to $Lie(N)$ (i.e. $Lie(B)=Lie(N)\stackrel{.}{+}V$). We can 
certainly pick a basis $t_i=(t_\iota,t_I)$ such that $t_\iota \in Lie(N)$ and $t_I \in V$ and complete
$(t_\iota,t_I)$ by the dual basis $(T^\iota,T^I)$ of $Lie(G)$.  From the duality property (5),  it follows 
that $T^\iota$'s span $V^\perp$ and $T^I$'s  span $Lie(N)^\perp$ (the superscript $\perp$ means
"perpendicular " in the sense of the bilinear form $(.,.)_\D$.) 
  We recall the formula (16b)
 $$w_{\L_R}(y)f= -\de_{t_i}(y)\nabla^R_{T^i}f= -\de_{t_\iota}(y)\nabla^R_{T^\iota}f -\de_{t_I}(y)\nabla^R_{T^I}f.$$
If  $y$ is  in $\rho^*(Fun(C))$, then $\delta_{t_\iota}(y)=0$ and we thus obtain
$$w_{\L_R}(y)f= -\de_{t_I}(y)\nabla^R_{T^I}f.$$
This means that $Lie(H)$ is spanned by  $T^I$'s only, or, in other words, $Lie(H)=Lie(N)^\perp$. 

 \medskip
 
\noindent Since  the twisting automorphism $\k$ preserves the
cosymmetry group $B$  the anomaly matrix $M_{\k^{-1}}^{ij}$ (cf. (13)) can be rewritten as
$$M_{\k^{-1}}^{ij}=(T^i,\k(T^m))_\D(\k(t_m), T^j)_\D=(P_\k T^i,T^j)_\D.\eqno(21)$$ Now we pick $u,v\in Fun(C)$ 
and, by using (12b)  and (21), we  calculate
$$\{\nu_R^*(u),\nu_R^*(v)\}_D= \{\L_R^*(\rho^*(u)),\L_R^*(\rho^*(v))\}_D=$$ $$=
\L_R^*\biggl(\{\rho^*(u),\rho^*(v)\}_B -M^{ab}_{\k^{-1}}\nabla^R_{t_a}\rho^*(u)\nabla^R_{t_b}\rho^*(v)\biggr)=$$
$$=
\L_R^*\biggl(\rho^*(\{u,v\}_C) -(P_\k T^A,T^B)_\D \nabla^R_{t_A}\rho^*(u)\nabla^R_{t_B}\rho^*(v)\biggr).$$
The transition  from the second to the third line is justified by the fact that $\nabla^R_{t_\al}\rho^*(u)=\nabla^R_{t_\bt}\rho^*(v)=0$ (Note that $a=(\al,A)$, $b=(\bt,B)$.) 
Since  both $T^A$'s and $T^B$'s are in $Lie(H)=Lie(N)^\perp$, we have  $(P_\k T^A,T^B)_\D =0$. Hence we conclude
that the moment map $\nu_R$ is non-anomalous:
$$\{\nu_R^*(u),\nu_R^*(v)\}_D=\nu_R^*(\{u,v\}_C).$$
\rightline{\#}

\noindent\underline{Remark}:  We have worked out the subsymmetry story for the right moment map $\L_R$. Obviously, there
is an analogous "left story" for which the conclusions are the same: a subgroup $H\subset G$
acting from the left (in the $\k$-twisted way) is the subsymmetry subgroup if $Lie(H) =Lie(N)^\perp$ where $Lie(N)$ is
the ideal in the cosymmetry Lie algebra $Lie(B)$. If, moreover, $P_\k(Lie(H))\subset Lie(N)$ then the 
$H$-subsymmetry is non-anomalous. We should also remark, that from
two conditions $[Lie(B),Lie(N)]\subset Lie(N)$ and $P_\k(Lie(H))\subset Lie(N)$ only
the second one is our  original result. The first one was already identified in \cite{ST1,FR}
for the
  \underline{non-twisted} Heisenberg doubles.   
  
  \subsection{Improper subsymmetry}

In this subsection,  we  partially release the condition of the decomposability of  twisted Heisenberg doubles in the
sense that we shall keep the unicity of the decomposition but not the globality.  Thus denote
$O_L$  the set of elements $K\in D$ for which it exists a  $g\in G$ and a   $b\in B$ such 
that $K=\k(b)g^{-1}$. In the same way, denote by $O_R$ the set of elements $K\in D$ for 
which it exists a  $\ti g\in G$ and a   $\ti b\in B$ such 
that $K=\k(\ti g)\ti b^{-1}$.  Suppose, moreover, that the respective  decompositions $\k(B)G$ and
$\k(G)B$  on $O_L$ and $O_R$ are unique. 

In the non-twisted case $\k=Id$, it was shown in \cite{AM} that  the lack of global decomposability has  unpleasant consequences. Namely, 
  the fundamental Semenov-Tian-Shansky Poisson structure (6) is no longer symplectic and, therefore, 
  the Poisson manifold $(D,\{.,.\}_D)$ cannot play the role of the phase-space of any dynamical system.
It turns out, however, that out from the Poisson structure $\{.,.\}_D$ one can construct symplectic submanifolds
of $D$ (called the symplectic leaves) which have the same dimension as $D$. In particular,  
Alekseev and Malkin have proved in \cite{AM} that the intersection $O_L\cap O_R$ is such 
symplectic leaf of $(D, \{.,.\}_D)$.   The result of Alekseev and Malkin can be generalized to the twisted
case as the following Lemma states:

\vskip1pc
\noindent\underline{Lemma 3}:

\noindent Let $(D,\k)$ be a twisted Heisenberg double and  $M$ its submanifold  defined as $M=O_L\cap O_R$. Consider
 maps $\L_L:M\to B$, $\Xi_R:M\to G$ induced by the unambiguous decomposition $M=\k(B)G$
 and maps $\Xi_L:M\to G$, $\L_R:M\to B$, induced by  $M=\k(G)B$ (thus $K=\k(\L_L(K))\Xi^{-1}_R(K)$ and  $K=\k(\Xi_L(K))\L_R^{-1}(K)$ for each $K$ in $M$). Denote by $r_G$ and $r_B$ the right-invariant Maurer-Cartan
 forms on $G$ and $B$, respectively (e.g. if $G$ is a matrix group $r_G=dgg^{-1}$). Then a two-form $\om_M$ on $M$ defined as
 $$\om_M=\jp(\L_L^*(r_B)\st \Xi^*_L(r_G))_\D +\jp (\L_R^*(r_B)\st \Xi^*_R(r_G))_\D\eqno(22)$$
 is  symplectic and its inverse is the fundamental  Poisson bivector (9) restricted to $M$.
 \vskip1pc
 \noindent\underline{Proof}:
 \noindent
 Choose a basis $t_i$ of $\B$ and $T^i$ of $\G$ fulfilling the duality relation $(T^i,t_j)_\D=\delta^i_j$.
 The form $\om_M$ can be then rewritten as  
 $$\om_M=\jp(\L_L^*(r_B),T^i)_\D\w( \Xi^*_L(r_G),t_i)_\D +\jp (\L_R^*(r_B),T^i)_\D\w( \Xi^*_R(r_G),t_i)_\D.$$
Denote by $<.,.>$ the pairing between forms
and vectors and recall the definition of the projectors $\Pi_{L\ti R},\Pi_{\ti LR},\Pi_{R\ti L}, \Pi_{\ti RL}$
from the proof of the Theorem 1.
 Then we have 
$$ <(\L_L^*(r_B),T^i)_\D, t>=(R_{K*}\k(T^i),\Pi_{L\ti R}t)_\D,\eqno(23a)$$
 $$ <(\Xi_L^*(r_G),t_i)_\D, t>=(R_{K*}\k(t_i),\Pi_{\ti L R}t)_\D,\eqno(23b)$$
 $$ <(\L_R^*(r_B),T^i)_\D, t>=-(L_{K*}T^i,\Pi_{R\ti L}t)_\D,\eqno(23c)$$
 $$ <(\Xi_R^*(r_G),t_i)_\D, t>=-(L_{K*}t_i,\Pi_{\ti R L}t)_\D,\eqno(23d)$$
 where  $t$ is a vector at a point $K$ of $M\subset D$.
 Let us show how to demonstrate (23abcd) on the example (23a). Due to the decomposability 
 $M=\k(B)G$, the vectors $L_{K*}T^i$, $R_{K*}\k(t_i)$ form the basis of the tangent space
 $T_KM$. Thus it is sufficient to prove (23a) for $t$ being one of the elements of the basis of $T_KM$.
For   $t=L_{K*}T^j$, it is obvious that the r.h.s. of (23a) vanishes. On the other hand, knowing that 
$\L_L(Ke^{sT^j})=\L_L(K)$, we can  evaluate  the l.h.s.:
$$ <(\L_L^*(r_B),T^i)_\D, L_{K*}T^j>=<(r_B,T^i)_\D,\L_{L*}(L_{K*}T^j)> =0.$$
For $t=R_{K*}\k(t_j)$, the r.h.s. of (23a) gives 
$$(R_{K*}\k(T^i),\Pi_{L\ti R} R_{K*}\k(t_i))_\D=(R_{K*}\k(T^i), R_{K*}\k(t_j))_\D=\delta^i_j.$$
On the other hand, knowing that $\L_L(e^{s\k(t_j)}K)=e^{st_j}\L_L(K)$, we can evaluate the l.h.s.:
$$ <(\L_L^*(r_B),T^i)_\D, R_{K*}\k(t_j)>=<(r_B,T^i)_\D,\L_{L*}(R_{K*}\k(t_j)> =$$ $$=<(r_B,T^i)_\D, R_{\L_L(K)*}t_j>= (R_{\L_L^{-1}(K)*}R_{\L_L(K)*}t_j,T^i)_\D=(t_j,T^i)_\D=\delta^i_j.$$
By using the relations (23abcd), we can evaluate the form $\om_M$ on   any two vectors $t,u\in T_KM$
in terms of the projectors:
$$\om_{M}(t,u)=\jp(R_{K*}\k(T^i),\Pi_{L\ti R}t)_\D(R_{K*}\k(t_i),\Pi_{\ti LR}u)_\D$$
$$- \jp(R_{K*}\k(T^i),\Pi_{L\ti R}u)_\D(R_{K*}\k(t_i),\Pi_{\ti LR}t)_\D$$
$$+\jp(L_{K*}T^i,\Pi_{R\ti L}t)_\D(L_{K*}t_i,\Pi_{\ti RL}u)_\D-\jp(L_{K*}T^i,\Pi_{R\ti L}u)_\D(L_{K*}t_i,\Pi_{\ti RL}t)_\D=$$
$$=\jp(\Pi_{L\ti R}t,\Pi_{\ti LR}u)_\D-\jp(\Pi_{L\ti R}u,\Pi_{\ti LR} t)_\D+\jp(\Pi_{R\ti L}t,\Pi_{\ti RL}u)_\D
-\jp(\Pi_{R\ti L}u,\Pi_{\ti RL}t)_\D.$$
By realizing that it holds
$$(t,\Pi_{\ti LR}u)_\D=(\Pi_{R\ti L}t,\Pi_{\ti LR}u)_\D=(\Pi_{R\ti L}t,u)_\D,$$
$$\Pi_{\ti LR}+\Pi_{R\ti L}=Id,$$
we  finally arrive at
$$\om_M(t,u)=(t,(\Pi_{\ti LR}-\Pi_{L\ti R})u)_\D.$$
From the equation (10), we know that the form $\om_M$ is invertible and its inverse is nothing
but the Semenov-Tian-Shansky Poisson tensor (9) restricted to $M$. From this
it also follows that $\om_M$ is closed hence symplectic.
\rightline{\#}
\noindent It is certainly a good news to have the symplectic submanifold $M$ of $D$, since it allows us
to construct dynamical systems also for globally non-decomposable twisted Heisenberg doubles.
On the other hand, it is a much less good news to remark that  nothing guarantees that the
group $G$ still acts on $M$. In fact, it turns out, generically, that  the submanifold $M$ of $D$
is not invariant under the left or right action of $G$ on $D$, therefore $G$ cannot play the role
of the symmetry group.  It may happen, however, that  there is a subgroup $H$ of $G$ which \underline{does}
 preserve the submanifold $M$ and which has the property that $\H=\N^\perp$, where $\N$ is an
 ideal in $\B$. We have then the following lemma
 
 \vskip1pc
 \noindent\underline{Lemma 4}: Let $H$ be a subgroup of $G$ preserving the submanifold $M=O_L\cap O_R$.
 We suppose moreover that $\H=\N^\perp$, where $\N$ is the ideal of $\B$. Then there exists a moment 
map $\nu:M\to B$
 realizing the global $(H,C)$-Poisson-Lie symmetry of $M$.
 \vskip1pc
 \noindent\underline{Proof}:
 For concreteness, we speak about the right action of $G$ on $D$. Sitting on $M$, we   construct the
 map $w_{\L_R}:Fun(B)\to Vect(M)$ by using the formula (3):
  $$w_{\L_R}(y)f=\{f,\L_R^*(y')\}_M \L_R^*(S(y'')), \quad y\in Fun(B), f\in Fun(M).$$ 
  For every $y\in Fun(B)$, we have obviously
  $$\nabla^L_{\k(T^i)}\L_R^*(y)=0.$$
  Since the  Poisson bivector on $M$ is given by Eq.(9), we thus obtain
  $$w_{\L_R}(y)f=\nabla^R_{T^i}f \nabla^R_{t_i}\L_R^*(y')\L_R^*(S(y''))=-\nabla^R_{T^i}f \L_R^*((\nabla^L_{t_i}y')S(y''))=
  -\delta_{t_i}(y)\nabla^R_{T^i}f.$$
  It follows that the Lie algebra $\G$ of $G$ does act $M$, however, because we have supposed it, this action cannot be lifted
  to the action of $G$ itself. Similarly as in the demonstration of  Theorem  3, we thus  observe that
  for $\nu_R\equiv \rho\circ \L_R$  the following is true
  $$\{f,\nu_R^*(u')\}_M \nu_R^*(S(u''))=-\delta_{t_I}(\rho^*(u))\nabla^R_{T^I}f , \quad u\in Fun(C), f\in Fun(M).$$
Recall that $T^I$'s span the Lie algebra $\H=\N^\perp$ therefore $\nu_R$ is indeed the moment map realizing the action of $\H$ on $M$. This action can be obviously lifted to the action of the group $H$ on $M$, since we have
supposed that $M$ is $H$-invariant. 

\rightline{\#}
\noindent\underline{Remark}: 

\noindent In the case of the non-decomposable Heisenberg doubles of the type just described we cannot
speak about the proper subsymmetry since $G$  does not act on $M$,  therefore we speak about the  improper subsymmetry.
 
\vskip1pc

\noindent Now it is time for an example. Consider a group $SL(3,R)$ (consisting of real $3\times 3$-matrices of unit determinant) and denote by $sl(3,R)$ its Lie algebra (consisting of real traceless $3\times 3$-matrices). 
The direct product $D=sl(3,R)\times SL(3,R)$ can be equipped with the group structure as follows:
$$(\chi,g)(\ti\chi,\ti g)=(\chi+Ad_g\ti\chi, g\ti g), \quad \chi,\ti\chi\in sl(3,R),\quad g,\ti g\in Sl(3,R),$$
$$(\chi,g)^{-1}=(-Ad_{g^{-1}}\chi,g^{-1}).$$
The Lie algebra $\D$ of $D$ is formed by pairs of elements of $sl(3,R)$ written as $\phi\oplus\al$
with the commutator
$$[\phi\oplus\al,\psi\oplus\beta]=  ([\phi,\beta] + [\al,\psi])\oplus [\al,\beta].$$
There is a natural bi-invariant metric on $D$ induced from  an invariant bilinear form $(.,.)_{\D}$  on $\D=Lie(D)$:
$$(\phi\oplus \al,\psi\oplus\beta)_\D= Tr(\phi\beta)+Tr(\psi\alpha), \quad \al,\beta,\phi,\psi\in sl(3,R).$$
The twisting automorphism $\k$ is defined by
$$\k(\chi,g)=(-\chi^T,(g^{-1})^T),$$
where $T$ stands for matrix transposition.
In order to establish that $(D,\k)$ is indeed a twisted Heisenberg double, we have to identify
two maximally isotropic subgroups. Here they are
$$G=\{(\chi,g)\in D; \chi =0\},$$
$$B=\biggl\{(\chi,g)\in D; \chi=\left(\begin{array}{ccc} \chi^\lhd+\chi^\rhd & \chi^{1+}&\chi^{3+}\\ \chi^{1-}&-2\chi^\rhd&\chi^{2+}\\ {1\over \e}(1- e^{-\e s})&\chi^{2-}&-\chi^\lhd+\chi^\rhd\end{array}\right),g=\left(\begin{array}{ccc} e^{\jp\e s} &0&-\e  e^{\jp\e s} \chi^\lhd\\ 0&1&0\\0&0&e^{-\jp\e s}  \end{array}\right)\biggr\},$$
where $s,\chi^\lhd,\chi^\rhd,\chi^{j+},\chi^{1-},\chi^{2-}\in\br $ are coordinates on $B$ and $\e$ is a parameter.
\vskip1pc
\noindent For the basis of $\D$, we may choose
$$T^\lhd=0\oplus H ,\ T^\rhd=0\oplus {K\over 3},\quad t_\lhd=2H\oplus (-\e E^{3+}) ,\ t_\rhd=2K\oplus 0,$$
$$T^{j+} =0\oplus E^{j+},\ T^{j-}=0\oplus E^{j-}, \quad t_{j+}= E^{j-}\oplus 0,\  t_{j-}= E^{j+}\oplus 0,\quad j=1,2, $$ 
$$T^{3+} =0\oplus E^{3+},\ T^{3-}=0\oplus E^{3-},\quad t_{3+} =E^{3-} \oplus \e H,\ t_{3-}=E^{3+}\oplus 0,$$
where 
$$E^{1+}=   \left(\begin{array}{ccc} 0&1& 0\\ 0&0&0\\0&0&0 \end{array}\right), \quad E^{2+}=   \left(\begin{array}{ccc} 0&0& 0\\ 0&0&1\\0&0&0 \end{array}\right),\quad E^{3+}=   \left(\begin{array}{ccc} 0&0& 1\\ 0&0&0\\0&0&0 \end{array}\right),$$
$$E^{1-}=   \left(\begin{array}{ccc} 0&0& 0\\ 1&0&0\\0&0&0 \end{array}\right), \quad E^{2-}=   \left(\begin{array}{ccc} 0&0& 0\\ 0&0&0\\0&1&0 \end{array}\right),\quad E^{3-}=   \left(\begin{array}{ccc} 0&0& 0\\ 0&0&0\\1&0&0 \end{array}\right),$$
 $$H=\left(\begin{array}{ccc} \jp&0&0 \\ 0&0&0\\0&0&-\jp \end{array}\right),\quad  K=\left(\begin{array}{ccc} \jp&0&0 \\ 0&-1&0\\0&0&\jp \end{array}\right) .$$
  It is easy to verify that it holds
$$(t_i,t_j)_\D=0,\quad (T^i,T^j)_\D=0,\quad (T^i,t_j)_\D=\delta^i_j,\quad i,j=\lhd,\rhd,1\pm,2\pm,3\pm.$$
The commutation relations of $\G=Span(T^i)$ are evidently those of the Lie algebra $sl(3,R)$. It is
important for us to give the complete list of (non-zero)  commutators of $\B=Span(t_i)$. Thus we have
$$[t_\lhd,t_{1+}]=\e t_{2-}, \ [t_\lhd, t_{2+}]=-\e t_{1-},\  [t_{3+},t_{3-}]=\e t_{3-}, \  [t_{3+},t_{\lhd}]=\e t_\lhd,$$
$$  [t_{3+},t_{j\pm}]=\mp \jp\e t_{j\pm}, \ j=1,2$$
Let us choose a (nilpotent) subalgebra $\H$ of $\G=sl(3,R)$ spanned by $T^{j+}$. Thus the only non-zero commutator is
$$[T^{1+},T^{2+}]=T^{3+}.$$
It is easy to find  $\N\subset \B$ such that $\H=\N^{\perp}$: we have
$$\N = Span(t_\lhd,t_\rhd, t_{j-}), \quad j=1,2,3.$$
It is the matter of direct check  to verify that $\N$ is indeed an ideal in $\B$.  Therefore the (Heisenberg) group $H$
consisting of upper-triangular real matrices with units on the diagonal is a good candidate for the Poisson-Lie subsymmetry.
The corresponding cosymmetry group $C$ has Lie algebra $\C=\B / \N$ and, by slightly abusing the notation, we can denote
its basis by $t_{j+}$, $j=1,2,3$. The non-zero commutators of $\C$ read
$$ [t_{3+},t_{j+}]=- \jp\e t_{j+}, \ j=1,2.$$
The cosymmetry group $C$ can be most easily described in the dual way. Denote the coordinate fonctions as $\xi^j$, $j=1,2,3$.
The coproduct reads
$$\Delta \xi^3=\xi^3\ot 1 +1\ot \xi^3, $$
$$\Delta \xi_j=\xi_j\ot 1+ e^{-{\e\over 2}\xi_3}\ot \xi_j,$$
the antipode
$$S(\xi_3)=-\xi_3,\ S(\xi_j)= -e^{{\e\over 2}\xi_3}\xi_j, \ j=1,2$$
and the counit
$$\ep(\xi_j)=0,\ j=1,2,3.$$
The dual map $\rho^*:Fun(C)\to Fun(B)$ reads
$$\rho^*(\xi_3)=s,\ \rho^*(\xi_j)=\chi^{j-}, \ j=1,2.$$
The Poisson-Lie bracket on $Fun(C)$ comes from that on $Fun(B)$, which, in turn, is given by (11).  The result of the computation reads
$$\{\xi^1,\xi^2\} _C={1\over \e}(1-e^{-\e\xi^3}),  \quad \{\xi^3,\xi^j\}_C=0,\ j=1,2.$$
We observe that both symmetry group  $H$ and the cosymmetry group  $C$ are non-Abelian.

\vskip1pc
\noindent  Let us now show that the $(H,C)$-Poisson-Lie subsymmetry is in fact improper. In order to see this, we first notice
  that the Heisenberg double $D$ is non-decomposable since e.g. the
element
$$(\chi,g)= \biggl(\left(\begin{array}{cc} 0 & -{1\over \e}\\0&0 \end{array}\right),\left(\begin{array}{cc} 1 & 0\\ 0&1 \end{array}\right)\biggr) $$  
cannot be written as $\k(b)g^{-1}$ for some $b\in B$ and $g\in G$.
\vskip1pc
\noindent It is easy to  identify the manifold $M=O_L\cap O_R$. We find
$$M=\biggl\{(\chi,g)\in D; \ Tr(J_L E^{3-})>-{1\over \e},\  Tr(J_R E^{3+})<{1\over \e}\biggr\},\eqno(24)$$
where we have  defined the $sl(3,R)$-valued functions $J_L,J_R$ on $D$ as
 $$J_L(\chi,g)=\chi,\quad J_R(\chi,g)=-Ad_{g^{-1}}\chi.$$
 The symplectic form on $M$ can be computed from the explicit expression (22). The result of   calculation is as follows
 $$\om_M= -\jp Tr(dJ_R\w l_G)+ \jp Tr(dJ_L\w r_G)+$$
$$- {\e\over 2} {Tr(dJ_L H)\w Tr(dJ_L E^{3-})\over 1+\e Tr(J_L E^{3-}) } -{\e\over 2} {Tr(dJ_R H)\w Tr(dJ_R E^{3+})\over 1-\e Tr(J_R E^{3+}) }.$$
Note that the left and right-invariant  Maurer-Cartan forms $l_G$, $r_G$ can be written also as $g^{-1}dg$, $dgg^{-1}$ since
$G=SL(3,R)$ is the matrix group. The explicit expression of the symplectic form $\om_M$ is quite illuminating in the sense that  it  explains why the constraints
$\ Tr(J_L E^{3-})>-{1\over \e},\  Tr(J_R E^{3+})<{1\over \e}$ in (24) had to be imposed. It is now the matter of direct inspection to find that the right   action of the group $H$ on $D$ and the left action of $\k(H)$ on $D$ preserve, respectively,  the symplectic manifold $M=O_L\cap O_R$.
The $(H,C)$-Poisson-Lie symmetry of $(M,\om_M)$ is therefore established.
\vskip1pc
\noindent

 \section{ $u$-deformed WZW model and its gauging}
 
 We begin this section by introducing a particular example of the deformation of the
 WZW model which was not discussed   in \cite{K,K1,K2}.
 Then we  shall perform the symplectic reduction of this $u$-deformed  WZW model with 
respect to a non-anomalous
quasi-adjoint action submoment map which is a sort of combination of the  moment maps
 constructed in Secs. 3.1 and 3.2. 
Finally, we shall argue why this quasi-adjoint symplectic reduction
can be interpreted as the gauging of the deformed WZW model.

\subsection{The $u$-deformation of the  WZW model}

It was conjectured in \cite{K} and explained in detail in  \cite{K2} that  the standard WZW model $\cite{W}$ on a compact Lie group $K$ is a dynamical system whose phase space can be identified with certain (decomposable) twisted Heisenberg double of a loop group $LK$.     Moreover,
 the symplectic form of the WZW model is just the inverse of the fundamental Semenov-Tian-Shansky  Poisson bivector (9). 
 The basic idea of the article \cite{K} can be rephrased as follows: since the loop group $LK$ may possess several different
 twisted Heisenberg doubles $(D,\k)$, it makes sense to consider the dynamical system based on each of  $(D,\k)$ as a sort
 of generalized WZW model. The (twisted Heisenberg) double of the standard WZW model is distinguished among all other
 doubles of the loop group $LK$ by the fact that  the cosymmetry group $B$ is Abelian. This circumstance is reflected by the
 fact that   the standard WZW model has the ordinary Hamiltonian symmetry structure.  On the other hand, the generalized WZW
 models have necessarily non-Abelian cosymmetry groups therefore their symmetry structure must be genuinely Poisson-Lie.
 Some  generalized WZW models form naturally  families parametrized by one or several parameters. Suppose we 
investigate such a family.  If for a particular value of the parameters the 
corresponding generalized WZW model becomes the standard WZW model, we call the other members of this 
family the \underline{deformed} WZW models.

\vskip1pc 
 \noindent Let us now describe a particular family of the  deformed WZW models, which was not discussed in \cite{K,K1,K2}.  Thus
 $K$ be a   connected simple compact   Lie group whose Lie algebra $\K$ is equipped with a non-degenerate $Ad$-invariant bilinear form $(.,.)_\K$ . Let $LK$ be the 
group of smooth maps from a circle $S^1$ into $K$ (the group law is given by pointwise
multiplication) and define a natural non-degenerate $Ad$-invariant bilinear form
 $(.\vert .)$ on $ L\K \equiv  Lie (LK)$ by the following formula
$$(\al\ve \bt)={1\over 2\pi}\int_{-\pi}^\pi d\si (\al(\si),\bt(\si))_{\K},   
 \eqno(25)$$
 As the twisted Heisenberg double $D$, we take the semidirect product
of the loop group $LK$ with its Lie algebra $L\K$.  Thus the group multiplication law on
 $D$  reads
$$(\chi, g).(\ti\chi, \ti g)=(\chi +Ad_g\ti\chi, g\ti g), \quad g\in LK, \chi\in L\K,\eqno(26a)$$
$$(\chi,g)^{-1}=(-Ad_{g^{-1}}\chi,g^{-1}),\eqno(26b)$$
and the  Lie algebra $\D$ of $D$ has the structure of semidirect 
 sum $\D =L\K\stackrel{\leftarrow}{ \oplus} L\K$ 
 $$[\phi\oplus \al,\psi\oplus \bt] = ([\phi,\bt]+[\al,\psi],[\al,\beta]).$$
 Here $\phi,\psi\in L\K$ are in the first and $\al,\bt\in L\K$ in the second composant of the semidirect sum. The bi-invariant metric on $D$ comes from $Ad$-invariant
 bilinear form $(.,.)_{\D}$ on $Lie(D)=\D$ defined with the help of (25):
 $$(\phi\oplus\al,\psi\oplus\bt)_{\D}=(\phi\ve \bt)+(\psi\ve \al). $$
  The  metric preserving automorphism $\k$ of  the group $D$ reads
 $$\k(\chi, g)= (\chi +k\d_{\si} gg^{-1},g),\eqno(26c)$$
 where $k$ is an (integer) parameter.  The maximally isotropic subgroups are 
$$G=\{(\chi,g)\in D; \chi =0\},\eqno(27a)$$
$$B=\{(\chi,g)\in D; g=e^{u(\chi)}\},\eqno(27b)$$
where $u$ is a certain map from $L\K$ to the Cartan subalgebra $\T$ of $L\K$. Let us 
now explain the construction of the map $u$:
 The group $K$ is naturally embedded in $LK$ as the subgroup consisting
of constant loops. The maximal torus $T$ of $K$ is therefore the (Abelian) subgroup 
of $LK$ and we call $\T=Lie(T)$ the Cartan subalgebra of $L\K$. Since we have the
inner product (25) on $L\K$ we can define  the orthogonal projector $\P_0:L\K\to \T$.
Let $U:\T\to\T$ be a skew-symmetric linear operator, i.e. it holds
$$(Ua,b)_\K=-(a, Ub)_\K, \quad a,b\in\T.\eqno(28)$$
We then define
$$u=U\circ \P_0.$$
It is easy to see that
$$u(\chi)+u(\ti \chi)=u\biggl(\chi+e^{u(\chi)}\ti\chi e^{-u(\chi)}\biggr),\quad \chi,\ti\chi\in L\K,$$
hence the set $B$  defined by (27b) is indeed the subgroup of $D$. Moreover,   the condition (28) implies the isotropy of  $B$
in $D$.

\vskip1pc
  \noindent It is a simple task to establish the decompositions $D=\k(G)B$ and $D=\k(B)G$.
 Indeed, we have for every
  $g\in LK,\chi\in L\K$
 $$(\chi, g) =  (k\d_\si gg^{-1},ge^{u(J_R)})(-e^{-u(J_R)}J_R e^{u(J_R)} ,
e^{-u(J_R)})=
(J_L,e^{u(J_L)}).(0,  e^{-u(J_L)}g), $$
 where  $L\K$-valued functions $J_L,J_R$ on $D$ are defined as
$$J_L(\chi,g)\equiv \chi, \quad J_R(\chi,g) = -Ad_{g^{-1}}\chi+kg^{-1}\d_\si g.\eqno(29a)$$ 
 Thus we can identify the moment maps $\L_{L,R}:D\to B$, $\Xi_{L,R}:D\to G$:
$$\L_L(\chi,g)= (J_L, e^{u(J_L)}),\quad 
\L_R(\chi,g)=(J_R,e^{u(J_R)}),\eqno(29b)$$
$$\Xi_L(\chi,g)=ge^{u(J_R)}, \quad \Xi_R(\chi,g)=g^{-1}e^{u(J_L)}.$$
Now we use the formula (22) and 
 write down the symplectic form $\om_u$ of the $u$-deformed 
 WZW model:
$$\om_u=\jp (dJ_L\w\vert r_{LK})-\jp(dJ_R\w\vert l_{LK}) + \jp (u(dJ_L)\w\vert dJ_L)
+\jp (u(dJ_R)\w\vert  dJ_R).\eqno(30)$$
Here $r_{LK}=dgg^{-1}$ and $l_{LK}=g^{-1}dg$ stand for the right and the left-invariant Maurer-Cartan forms on the group manifold $LK$. 
\vskip1pc
 \noindent  The role of the deformation parameter is played by the linear operator $U$. Indeed,  if $U\to 0$ the form $\om_u$
 can be rewritten as
 $$\om_{u=0}=d(J_L\vert r_{LK})+\jp k(r_{LG}\w \vert \d_\si r_{LG} ) .$$
 In the expression $\om_{u=0}$, we can recognize the symplectic form of the standard WZW model (cf. \cite{K,Gaw,Mad}).
 We now complete the definition of the $u$-deformed WZW model by saying that it is a dynamical system with the phase space $D$,
 with the symplectic form $\om_u$ and with the following  Hamiltonian
 $$H=-{1\over 2k}(J_L\vert J_L) -{1\over 2k}(J_R\vert J_R).\eqno(31)$$
 We note without giving proof that, in distinction to the $q$-deformation of the WZW model introduced
 in \cite{K},  the $u$-deformation does preserve the conformal symmetry.
 \vskip1pc
 \noindent   Let us study the symmetry structure of the $u$-WZW model. The group $G=LK$ acts from the left  as
 $$h\rhd (\chi,g)= \k((0,h)).(\chi, g)= (k\d_\si hh^{-1} +h\chi h^{-1},hg), \quad h,g\in LK, \ \chi\in L\K$$
 and also from the right
 $$(\chi,g)\lhd h= (\chi,g)(0,h^{-1})=(\chi,gh^{-1}).$$
 We know (by construction) that both these actions are Poisson-Lie symmetries with the moment maps $\L_{L,R}$ given by (29b).
 Now we are going to evaluate the (anomalous) Poisson brackets (12ab) of the moment maps. First of all,  we have to describe the
 structure of the cosymmetry group $B$ in the dual language. The complexified algebra  $Fun^\bc(B)$
  is generated by (linear) functions 
$F^{\al, n}, F^{\mu, n}$  
  defined as
  $$F^{\al,n}(\chi) = (E^{\al,n}\ve \chi), \quad  F^{\mu,n}(\chi) = (H^{\mu,n} \ve \chi), \quad \chi\in L\K .\eqno(32)$$
  Here $E^{\al,n}=E^\al e^{in\si}$ and $E^\al$ are the step generators of the complexified Lie algebra $\K^\bc$. On the other hand, $H^{\mu,n}=H^\mu e^{in\si}$ where $H^\mu$ are the  (orthonormalized) Cartan generators fulfilling the
  relations
  $$[H^\mu,E^\al]=<\al,H^\mu>E^\al,\quad  [E^\al,E^{-\al}]=\al^\vee ,\quad [E^\al,E^\bt]=c^{\al\bt}E^{\al+\bt},$$
  $$ (H^\mu,H^\nu)_\K=\delta^{\mu\nu}, \quad (E^{\al},E^{-\al})_{\K^\bc}={2\over \ve \al\ve ^2}, \quad (E^\al)^\dagger=E^{-\al},\quad (H^\mu)^\dagger = H^\mu,$$
  where the coroot $\al^\vee$ is defined as
  $$\al^\vee ={2\over \ve \al\ve ^2}<\al,H^\mu>H^\mu.$$
  Obviously, $E^{\al,n}, H^{\mu,n} $, $n\in \bz$ is the basis of $L\K^\bc$.
 The (non-Abelian)
 group law on $B$ is   encoded in the coproduct, the antipode and the counit on $Fun^\bc(B)$.  From the Eqs. (26), (27b) and (32), it is 
 not difficult to find out: 
  $$\Delta F^{\mu, n}=F^{\mu, n}\otimes 1+1\otimes F^{\mu, n}, \quad S(F^{\mu, n})=-F^{\mu, n}, \quad \e(F^{\mu, n})=0, \quad \e(F^{\al, n})=0,$$ 
    $$\Delta F^{\al, n}=F^{\al, n}\otimes 1 +  e^{-<\al ,U(H^\mu)>F^{\mu,0}}\otimes F^{\al, n} , 
    \quad S(F^{\al, n})=-e^{<\al,U(H^\mu)>F^{\mu,0}}F^{\al, n}.$$ 
Because of the fact that $\chi^\dagger =-\chi$, the operation of the complex conjugation $\dagger$ on $Fun^\bc(B)$ is given by
   $$(F^{\al,n})^\dagger=-F^{-\al,-n},\quad (F^{\mu,n})^\dagger=-F^{\mu,-n}.$$ It can be then easily verified that 
    $$\Delta \circ \dagger =(\dagger\ot\dagger)\circ \Delta  , \quad S\circ \dagger=\dagger\circ S,\quad \e \circ \dagger=\dagger\circ \e.$$
    This means that  $\Delta,S,\e$ descend from $Fun^\bc (B)$ to $Fun^\br (B)$ making the latter the real commutative Hopf algebra
   dual to  the real group $B$.
  \vskip1pc 

 \noindent  The Poisson-Lie bracket  on $Fun^\bc(B)$ can be obtained from the general formula (11):
  $$\{F^{\mu,m}, F^{\nu,n}\}_B=0,$$
    $$\{F^{\mu,m}, F^{\al,n}\}_B=<\al,H^\mu> F^{\al,m+n},$$
        $$\{F^{\al,m}, F^{-\al,n}\}_B={2\over \ve \al\ve ^2}<\al,H^\mu>F^{\mu,m+n},$$
                $$\{F^{\al,m}, F^{\bt,n}\}_B=c^{\al\bt} F^{\al+\bt,m+n}-<\al,U(H^\mu)><\bt, H^\mu>   F^{\al,m}F^{\bt,n}.$$
                It is easy to verify, that  the Poisson-Lie bracket on $Fun^\bc(B)$ verifies
                $$\{f_1^\dagger,f_2^\dagger \}_B=\{f_1,f_2\}_B^\dagger,$$
                hence it defines also the Poisson-Lie bracket on the real group $B$. 
            Now we are ready to evaluate the anomalous Poisson brackets (12ab). We start with
  
 $$\L_L^*(F^{\al,n})= (J_L\ve E^\al e^{in\si})\equiv J_{L}^{\al,n},\quad \L_L^*(F^{\mu,n})= (J_L\ve H^\mu e^{in\si})\equiv J_{L}^{\mu,n}, $$
   $$\L_R^*(F^{\al,n})= (J_R\ve E^\al e^{in\si})\equiv J_{R}^{\al,n},\quad \L_R^*(F^{\mu,n})= (J_R\ve H^\mu e^{in\si})\equiv J_{R}^{\mu,n}$$  
   and find
   
  $$\{J_L^{\mu,m}, J_L^{\nu,n}\}_D= {k\delta^{\mu\nu}in\delta_{m+n,0}},$$
    $$\{J_L^{\mu,m}, J_L^{\al,n}\}_D=<\al,H^\mu> J_L^{\al,n+m},$$
        $$\{J_L^{\al,m}, J_L^{-\al,n}\}_D={2\over \ve \al\ve ^2} \biggl(<\al,H^\mu>J_L^{\mu,n+m} {+ikn\delta_{m+n,0}}\biggr),$$
                $$\{J_L^{\al,m}, J_L^{\bt,n}\}_D=c^{\al\bt}J_L^{\al+\bt,m+n}\underline{  -<\al,U(H^\mu)><\bt, H^\mu>   J_L^{\al,m}J_L^{\bt,n}};\eqno(33a)$$ 
                
    $$\{J_R^{\mu,m}, J_R^{\nu,n}\}_D= -{k\delta^{\mu\nu}in\delta_{m+n,0}},$$
    $$\{J_R^{\mu,m}, J_R^{\al,n}\}_D=<\al,H^\mu> J_R^{\al,n+m},$$
        $$\{J_R^{\al,m}, J_R^{-\al,n}\}_D={2\over \ve \al\ve ^2} \biggl(<\al,H^\mu>J_R^{\mu,n+m} {-ikn\delta_{m+n,0}}\biggr),$$
                $$\{J_R^{\al,m}, J_R^{\bt,n}\}_D=c^{\al\bt}J_R^{\al+\bt,m+n}\underline{  -<\al,U(H^\mu)><\bt, H^\mu>   J_R^{\al,m}J_R^{\bt,n}};\eqno(33b)$$ 
               
                $$\{J_L,J_R\}_D=0.\eqno(33c)$$
 In the formulae above, we note  the anomalous terms proportional to $k$. They correspond to the 
matrices $M^{ij}_\k$ and $M^{ij}_{\k^{-1}}$ in (12a) and (12b), respectively. We remark, that 
the left and right brackets differ by the sign in front of $k$. This fact will be crucial
for gauging the $u$-deformed WZW model in Sec 4.3.  We have also underlined the defomation terms
containing $U$. Thus the relations (33a) or (33b) can be referred to as those of $u$-deformed Kac-Moody algebra.
                               
 \vskip1pc
\noindent Knowing    the symplectic structure of the $u$-deformed
WZW models, we can compute  other interesting Poisson brackets.  The observables
on $D$ are functions of $\chi\in L\K$ and $g\in LK$. Let as consider two  functions $\phi(g),\psi(g)$, 
which do not depend on $\chi$. Then we find directly from (6):
$$\{\phi(g),\psi(g)\}_D=\underline{\nabla^R_{T^\mu}\phi(g) \nabla^R_{U(T^\mu)}\psi(g)
-\nabla^L_{U(T^\mu)}\phi(g)\nabla^L_{T^\mu}\psi(g)},$$
where $T^\mu\equiv iH^\mu\in \T\subset\K$.
Note, that we have again underlined the $u$-deformation term (the corresponding bracket of the
standard WZW model vanishes). Finally, we
have 
$$\{\phi(g),J_L^{\mu,m}\}_D=\nabla^L_{H^{\mu,m}}\phi(g),$$
$$\{\phi(g),J_L^{\al,n}\}_D= \nabla^L_{E^{\al,n}}\phi(g) \underline{-i<\al,U(H^\mu)>J_L^{\al,n}\nabla^L_{T^{\mu}}\phi(g)},$$
$$\{\phi(g),J_R^{\mu,m}\}_D=-\nabla^R_{H^{\mu,m}}\phi(g),$$
$$\{\phi(g),J_R^{\al,n}\}_D=- \nabla^R_{E^{\al,n}}\phi(g) \underline{+i<\al,U(H^\mu)>J_R^{\al,n}\nabla^R_{T^{\mu}}\phi(g)},$$

  \subsection{Symplectic reduction: generalities}

 The symplectic reduction is the method of  construction of new symplectic manifolds out from old ones.  
 The simplest way
 of explaining  the method  relies on the dual language which uses  rather
 the algebra of  functions $Fun(M)$ on a symplectic manifold $M$  than   the manifold $M$ itself.   We note that the space $Fun(M)$ is the 
 Poisson algebra, i.e. the Lie algebra compatible with the structure of the (standard commutative point-wise) multiplication on $Fun(M)$.
 The Lie commutator is nothing but the Poisson bracket  
 $\{.,.\}_M$ corresponding to a symplectic structure $\om_M$ on $M$ and the compatibility condition is given by the
 Leibniz rule:
 $$\{f,gh\}_M=\{f,g\}_Mh+\{f,h\}_Mg,\quad f,g,h\in Fun(M).$$
  Let $J$ be an ideal of the algebra $Fun(M)$ with
 respect  to the ordinary commutative multiplication on $Fun(M)$ (typically, $J$ is the ideal of functions vanishing on a submanifold
 $N\subset M$).  Let $J$ be also the Poisson subalgebra of $Fun(M)$, i.e. $\{J,J\}\subset J$.  We can now construct a new Poisson algebra $\ti A$  defined as follows
 $$\ti A =\{f\in Fun(M); \ \{f,J\}_M\in J\}.$$
 Note that the property $\{J,J\}\subset J$ implies that $J\subset \ti A$.
 By construction,  $J$ is not only the ordinary ideal of $\ti A$ but it is also the Poisson ideal, i.e. $\{\ti A,J\}_M\subset J$.  Obviously,
 the factor algebra $A_r\equiv \ti A/ J$  inherits the Poisson bracket from $\ti A$ hence it becomes itself the Poisson algebra.
  If $J$ is the ideal of functions vanishing on a submanifold $N\subset M$, then the   algebra $A_r$ is nothing but the
  Poisson algebra of functions corresponding to  some symplectic  manifold $M_r$. The manifold $M_r$ together with its corresponding Poisson bracket $\{.,.\}_r$ (or, equivalently, with its symplectic form $\om_r$)
    is called the reduced symplectic manifold. If there is a Hamiltonian $H$ on $M$  such that $H\in \ti A$, its class in $\ti A/J$ is denoted as $H_r$ ant it is referred to as the reduced Hamiltonian. 
  \vskip1pc
 \noindent The symplectic reduction is often put in relation with the actions of  Lie groups on the non-reduced manifold $M$. It may even happen that the reader used to the group approach   to  the symplectic
 reduction did not recognize at first reading that  his way of thinking about the reduction is just a particular case
 of the general  algebraic definition presented above.  We believe that it is worth to elucidate this point not only for pedagogical reasons.
 In fact, the group-based symplectic reduction will turn out to be in the core of our gauging of the 
 $u$-WZW model.  We shall work in
 the general Poisson-Lie setting, the standard Hamiltonian symplectic reduction (cf. \cite{OR} and references therein) will be the special case of our discussion
 when the cosymmetry group $B$ is Abelian. 
 \vskip1pc
 \noindent
 Suppose that there is a \underline{non-anomalous} moment map $\mu:M\to B$ realizing the $(G,B)$-Poisson-Lie symmetry
 of $M$ (cf. the Definition 1 of Section 2.2). Due to the property (2b) of the Poisson-Lie bracket on $Fun(B)$, we know that 
 the kernel of the counit $Ker(\ep)$ is the Poisson subalgebra of $(Fun(B),\{.,.\}_B)$.  Since the moment map $\mu$ is non-anomalous,
 the pull-back $\mu^*(Ker(\ep))$  is also the Poisson subalgebra of $(Fun(M),\{.,.\}_M)$. Thus the role of the ideal $J$ from the
 general definition above is played by the ideal of $Fun(M)$ generated by $\mu^*(Ker(\ep))$.We denote it also by the letter $J$.
 In the situation just described, the resulting reduced symplectic manifold $M_r$ (corresponding to the reduced Poisson algebra $\ti A/J$), can be easily "visualised". For this, let us suppose that the set $P$ of points of $M$ mapped by $\mu$ to the unit element
 $e$ of the cosymmetry group $B$ forms a smooth submanifold of $M$. It is not difficult to verify that the action of the symmetry
 group $G$ (which is itself locally induced by the moment map $\mu$) leaves $P$ invariant. Let us moreover suppose that the
 $G$-action on $P$ is free, or, in other words, that $P$ is isomorphic to a principal $G$-bundle. Then the basis $P/G$ of this $G$-fibration
can be then identified with the reduced symplectic manifold $M_r$.  The restriction of the symplectic
form $\om$ on $P$ becomes degenerated and the degeneracy direction of $\om$ turn out to be nothing but the
orbits of the gauge group  $G$. Thus the  symplectic form $\om_r$  is naturally induced from $\om$. Indeed, on each local trivialisation
of the $G$-bundle $P$ we can choose a slice. The restriction of $\om$ on the slice is the reduced symplectic form $\om_r$. 

\vskip1pc
\noindent A particularly good situation occurs when the $G$ fibration of $P$ is topologically
trivial. In this case,  one can visualize the reduced symplectic manifold as the submanifold
of $P$ (and, hence, as the submanifold of the original symplectic manifold $M$). This can be
done by choosing a global slice $Q_i=0$, where the functions $Q_i$ are in $Fun(M)$. In the
usual terminology, the functions  $J_i\in\mu^*(Ker(\ep))\subset Fun(M)$ are called the first class
constraints and the functions $Q_i$ their complementary  second class constraints. The reduced symplectic
manifold $M_r$ is now the common locus of   all   constraints $J_i=0$ and $Q_i=0$ and
the reduced symplectic form $\om_r$ is the pull-back of the non-reduced form $\om$ to the submanifold $M_r$.
\vskip1pc
\noindent It is sometimes convenient to fix the gauge only partially. This means that it exists a  slice
$Q_\ga=0$ (the subscript $\ga$ runs over a smaller set than the subscript $i$) which restricts the gauge freedom to some subgroup $H\subset G$. If we note by the letter $L$
the common locus $J_i=0,Q_\ga=0$ in $M$, the reduced symplectic manifold $M_r$ can be identified with the
coset space $L/H$. The interest in such partial gauge fixing will be  evident in the studies of the symplectic structure
of the standard gauged WZW model and of its deformations. Indeed,  as we shall see in the following section,   there exists the partial gauge fixing for which the manifold $L$ has 
a very simple left-right chiral  symmetric description and the  residual gauge group $H$ is finite
dimensional, compact and Abelian.   

\subsection{Symplectic reduction of the $u$-WZW model}
We start this section by remarking that the twisting automorphism $\k$ given by (26c) not only preserves
the cosymmetry group $B$ described in (27b) but it leaves invariant every element of $B$.  This means that we can safely apply the Theorem 2 of Sec. 3.1. which now states that the products $ \L_L \L_R \equiv B_L$ and $\L_R \L_L\equiv B_R$ are both \underline{non-anomalous} moment maps.   We already know from the general theory that  both $B_L$ and $B_R$ realize
the global Poisson-Lie symmetries of the twisted Heisenberg double $(D,\k)$ therefore, via their corresponding maps $w_{B_L}$,$w_{B_R}$  (cf. (3)), they induce the respective actions (20a),(20b)  of the loop group $G=LK$ on $(D,\k)$.  

Let us work, for concreteness, with the moment map $B_L=\L_L\L_R$. Recall the group multiplication law in $B$:
$$(\chi_1,e^{u(\chi_1)}).(\chi_2,e^{u(\chi_2)})=(\chi_1+e^{u(\chi_1)}\chi_2 e^{-u(\chi_1)}, e^{u(\chi_1)+u(\chi_2)}),\quad \chi_1,\chi_2\in L\K.\eqno(34)$$
The formula (34) together with Eqs. (29b) allow us to calculate the $B_{L,R}^*$-pull-backs of the basic functions from $Fun^\bc(B)$:
$$B^*_L(F^{\al,n})=(\L_L\L_R)^*(F^{\al,n})=J_L^{\al,n}+e^{-<\al,U(H^\mu)J_L^{\mu,0}}J_R^{\al,n},$$
$$B^*_R(F^{\al,n})=(\L_R\L_L)^*(F^{\al,n})=J_R^{\al,n}+e^{-<\al,U(H^\mu)J_R^{\mu,0}}J_L^{\al,n},$$
$$B^*_L(F^{\mu,n})=B^*_R(F^{\mu,n}) =J_L^{\mu,n}+ J_R^{\mu,n}.$$
Now we are ready to make explicit the  map $w_{B_L}: Fun(B)\to Vect(D)$:
$$w_{B_L}(F^{\al,n})f\equiv \{f,B_L^*((F^{\al,n})')\}_D B_L^*(S((F^{\al,n})''))=$$
$$=\nabla^L_{\k(E^{\al,n})}f-e^{-<\al,U(H^\mu)>J_L^{\mu,0}}\nabla^R_{E^{\al,n}}f-<\al,U(H^\mu)>J_L^{\al,n}\nabla^R_{H^\mu}f,$$
$$w_{B_L}(F^{\mu,n})f\equiv \{f,B_L^*((F^{\mu,n})')\}_D B_L^*(S((F^{\mu,n})''))=$$
$$=\nabla^L_{\k(H^{\mu,n})}f- \nabla^R_{H^{\mu,n}}f, \quad f\in Fun^\bc(D). $$
Recall that the symbol $w_{B_L}(F^{\al,n})$ denotes the (complex) vector field on $D$ corresponding to the Poisson-Lie 
Hamiltonian $F^{\al,n}\in Fun^\bc(B)$. Similarly, we find
$$w_{B_R}(F^{\al,n})f\equiv \{f,B_R^*((F^{\al,n})')\}_D B_R^*(S((F^{\al,n})''))=$$
$$=-\nabla^R_{E^{\al,n}}f+e^{-<\al,U(H^\mu)>J_R^{\mu,0}}\nabla^L_{\k(E^{\al,n})}f+<\al,U(H^\mu)>J_R^{\al,n}\nabla^L_{H^\mu}f,$$
$$w_{B_R}(F^{\mu,n})f\equiv \{f,B_R^*((F^{\mu,n})')\}_D B_R^*(S((F^{\mu,n})''))=$$
$$=\nabla^L_{\k(H^{\mu,n})}f- \nabla^R_{H^{\mu,n}}f, \quad f\in Fun^\bc(D). $$
\vskip1pc
\noindent It is the matter of easy check that the vector fields  $w_{B_L}(F^{\al,n})$,$w_{B_L}(F^{\mu,n})$ 
and also  $w_{B_R}(F^{\al,n})$,$w_{B_R}(F^{\mu,n})$ generate the actions of
the Lie algebra $L\K^\bc$ on $Fun^\bc(D)$. Moreover, it can be also seen that, by considering only the
Poisson-Lie Hamiltonians from $Fun^\br(B)$, these actions get restricted to the actions of $L\K$ on $Fun^\br(D)$. 
It is not difficult to lift the $L\K$ actions just described to the $LK$ actions. The resulting formulae are the special cases of the general formulae  (20a) and (20b):
$$h\tr (\chi,g)= \k(h)(\chi,g)h^{-1}_L,  \quad h_L=e^{-u(hJ_Lh^{-1}+\k \partial hh^{-1})}he^{u(J_L)},\quad h\in LK, \eqno(35a)$$
 $$h\tr (\chi,g)= \k(h_R)(\chi,g)h^{-1},  \quad h_R=e^{-u(hJ_Rh^{-1}-\k \partial hh^{-1})}he^{u(J_R)}, \quad h\in LK.\eqno(35b)$$
 We notice that for $U\to 0$ the cosymmetry group $B$ becomes Abelian and the $LK$-actions
 (35a) and (35b) coincide and (as we have promised to show in Section 3.1) they become identical to the twisted adjoint action $h\tr (\chi,g)=\k(h)(\chi,g)h^{-1}$.
 \vskip1pc
 \noindent  Let $\U$ be a subset of the set of all positive roots of the Lie algebra $\K^\bc$. Consider  a complex vector space $\S^\bc$ defined as
 $$\S^\bc=Span\{E^\ga,E^{-\ga}, [E^\ga,E^{-\ga}]\}, \quad \ga\in\U.$$
 In the rest of this paper, we shall suppose that the subset $\U$ was chosen in such a way that
 the vector space $\S^\bc$ is the Lie subalgebra of $\K^\bc$ (as an example take the block diagonal embedding
 of $sl_3$ in $sl_4$). Obviously, the vector space
 $$\T_S^\bc=Span\{[E^\ga,E^{-\ga}] \},\ \ga\in \U$$ is the Cartan subalgebra of $\S^\bc$.  
 The complex Lie algebra $\S^\bc$ has a natural compact real form $\S$ consisting
 of the anti-Hermitean elements of $\S^\bc$.  Consider the corresponding compact semi-simple group
 $S$ and view it as the  subgroup of $K$.    We are now going to establish the conditions on the operator $U$ which will  guarantee  that the action of the loop group $LS$ on $D$ via (35a) or (35b) is the Poisson-Lie subsymmetry.  
 \vskip1pc
 \noindent  Suppose that for all $\ga\in\U$, the operator $U:\T\to \T$ fulfils the following condition
 $$(\ga\circ U)(\T_S^\perp)=0,\eqno(36)$$
where the subscript $\perp$ stands for the orthogonal complement with respect to the
restriction of   the Killing-Cartan form $(.,.)_\K$ to $\T$. It is then easy to verify  that the set
$$N=\{(\chi,g)\in D; \ g=e^{u(\chi)}, \chi\in \S^\perp\}$$
is the normal subgroup of $B$. Consider  the algebra of complex functions on the group
$C=B/N$. As we have learned in Section 3.2,
$Fun^\bc(C)$ can be injected by the map $\rho^*$ into $Fun^\bc(B)$. (Note that $\rho^*$ is the
dual map  to the projection homomorphism $\rho: B\to B/C$.) It is
easy to see that $\rho^*(Fun^\bc(C))$ is spanned by the functions $F^{\ga,n},F^{\nu,n}$ where
$\ga\in \U$ and $H^\nu\in \T_S$.  The normality of the subgroup $N$ implies that
the vector space $\rho^*(Fun^\bc(C))$ is in fact the Hopf subalgebra of $Fun^\bc(B)$. By using the
explicit form of the Poisson-Lie brackets on $Fun^\bc(B)$,  it is straightforward
to check that $\rho^*(Fun^\bc(C))$ is also the Poisson subalgebra of $Fun^\bc(B)$.  It is 
moreover true that  $\rho^*(Fun^\bc(C))$ is $\dagger$-invariant hence we conclude that
$\rho^*(Fun(C)$ is the Poisson subalgebra of $Fun(B)$. All that  means that we can use
the Theorem 3 of Section 3.2 to conclude that  the action of the loop group $LS$  on $D$
via (35ab) is the Poisson-Lie subsymmetry. Our next goal is to gauge this (non-anomalous) subsymmetry, or, in other
words, to perform the symplectic reduction with respect to it.
\vskip1pc
\noindent  Consider the $LS$-subsymmetry moment map $C_L=\rho\circ B_L$, where $\rho$ 
is the projection homomorphism from $B$ to $C=B/N$.  The first step of the reduction procedure
consists in identification of the submanifold $P_L\subset D$ such that every point $p\in P_L$
is mapped by $C_L$ to the unit element of the group $C$. It is easy to see that
$$P_L=\{p\in D; \ J_L^{\ga,n}(p)+e^{-<\ga,U(H^\nu)>J_L^{\nu,0}(p)}J_R^{\ga,n}(p)=0,\ J_L^{\nu,n}(p)+J_L^{\nu,n}(p)=0\},$$
where $\ga\in \pm\U$ and $\nu$ is such that $H^\nu\in \T_S$. In physicists' terminology, the expressions $$J_L^{\ga,n} +e^{-<\ga,U(H^\nu)>J_L^{\nu,0}}J_R^{\ga,n}=0,\ J_L^{\nu,n}+J_L^{\nu,n}=0\eqno(37)$$ are the first class
constraints since it is not difficult to verify that the Poisson brackets of the constraints
among themselves as well as those of the Hamiltonian (31) with the
constraints vanish on the constrained surface $P_L$.

\vskip1pc
\noindent   Now the $u$-deformed WZW symplectic form $\om_u$ restricted to $P_L$ becomes degenerated in the directions of the action of $LS$ on $P_L$.  As we already know  from  Section 4.2, the reduced symplectic manifold
$M_r$ can be identified with the coset space $P_L/LS$.   We now perform a partial gauge fixing (cf. the general
discussion in Section 4.2) which will lead to very elegant left-right symmetric  chiral description of the symplectic
structure of the reduced symplectic manifold $M_r$.  For this, we first study the action of $LS$ on $D$ given by the
formula (35a). By using the formula (7a), we rewrite it as follows
$$s\tr (\chi, g)=(s\chi s^{-1} +k \partial_\si ss^{-1}, sgs_L^{-1}), \quad s_L=e^{-u(sJ_Ls^{-1}+\k \partial ss^{-1})}se^{u(J_L)},\  s\in LK . \eqno(38)$$
It is convenient to decompose $\chi$ as $\chi_s +\chi_p$, where $\chi_s\in L\S$ ans $\chi_p\in L\S^{\perp}$. We thus see from Eq. (38) that  
$\chi_s$ and $\chi_p$ do not mix under the action of $s$.  We know  that  every $\chi_s$ can be brought by some
$s$ to an element of   the finite dimensional Cartan subalgebra $\T_S$  (cf. \cite{K}, Theorem 3.6). Having in mind
the definition (29a) of $J_L$, this leads to the following natural slice on $D$:
$$J_L^{\ga,n}=0, \quad \ga\in\pm\U, \ n\in \bz,\eqno(39a)$$
$$J_L^{\nu,n}=0, \quad n\in \bz, \ n\neq 0,\eqno(39b)$$
where $\nu$ is such that $H^\nu\in \T_S$. This   slice is partial (it corresponds to the slice $Q_\ga=0$ in the general discussion of Sec. 4.2). Indeed,
the residual gauge group $H$   is the normalizer of the Cartan subalgebra $\T_S$ and, as
the discussion before the Theorem 3.6 of \cite{K} implies, the finite-dimensional Cartan  torus 
$T_S$ is the normal subgroup of $H$. (In fact $H/T_S$ is nothing but the affine Weyl group
of $LS$).     The constraints (37) and (39) can
be now rewritten in a $U$-independent way as
$$J_L^{\ga,n}=0, \quad J_R^{\ga,n}=0,\quad \ga\in\pm\U, \ n\in \bz,\eqno(40a)$$
$$J_L^{\nu,n}=0,\quad J_R^{\nu,n}=0, \quad  n\in \bz, \ n\neq 0.\eqno(40b)$$
$$J_L^{\nu,0}+J_R^{\nu,0}=0,\eqno(40c)$$
where $\nu$ is such that $H^\nu\in \T_S$. The constraints (40) define the submanifold $L\subset D$ and
the reduced symplectic manifold $M_r$ can be identified with the space of cosets $L/H$. 
\vskip1pc
\noindent The similar discussion can be performed also with the moment map $C_R=\rho\circ B_R$.
The first class constrained manifold $P_R$ is
$$P_R=\{p\in D; \ J_R^{\ga,n}(p)+e^{-<\ga,U(H^\nu)>J_R^{\nu,0}(p)}J_L^{\ga,n}(p)=0,\ J_L^{\nu,n}(p)+J_L^{\nu,n}(p)=0\},\eqno(41)$$
where $n\in \bz$, $\ga\in \pm\U$ and $\nu$ is such that $H^\nu\in \T_S$. The partial slice on $D$ is
$$J_R^{\ga,n}=0, \quad \ga\in\pm\U, \ n\in \bz,\eqno(42a)$$
$$J_R^{\nu,n}=0, \quad n\in \bz, \ n\neq 0,\eqno(42b)$$
where $\nu$ is such that $H^\nu\in \T_S$. 
The constrains (41) and (42) can also be rewritten in the $U$-independent way as
$$J_L^{\ga,n}=0, \quad J_R^{\ga,n}=0,\quad \ga\in\pm\U, \ n\in \bz,\eqno(43a)$$
$$J_L^{\nu,n}=0,\quad J_R^{\nu,n}=0, \quad  n\in \bz, \ n\neq 0.\eqno(43b)$$
$$J_L^{\nu,0}+J_R^{\nu,0}=0,\eqno(43c)$$
We thus see  that the symplectic reduction based on the moment map $B_R$ gives the
same result as the one based on $B_L$. This happens inspite of the fact that $w_{C_L}$ and
$w_{C_R}$ induce the different actions   of the gauge group $LS$ on $D$.
\vskip1pc
\noindent  Our next task will be the description of the symplectic form $\om_r$ on $M_r$. Actually, we shall
describe the pull-back of the original Semenov-Tian-Shansky form $\om_u$ on $D$ to the submanifold
$L\subset D$.  We again use the Theorem 3.6 of \cite{K} which permits us to parametrize the Heisenberg
double $D$ by means of two elements $g_L,g_R$ of $LK$ and one element $\mu$ of the Weyl alcove
$\A_K$ in the Cartan subalgebra $\T_K\subset \K$:
$$(\chi, g)= \k(0,g_L)(\mu,e_{LK})(0,g_R)^{-1}=(g_L\mu g_L^{-1} +k\partial_\si g_L g_L^{-1}, g_Lg_R^{-1}).\eqno(44)$$
Here $e_{LK}$ is the unit element in $LK$. 
 The Semenov-Tian-Shansky form $\om_u$ given by (30) gets rewritten in the new variables as follows
$$\ti\om_u=-d(\mu\ve g_R^{-1}dg_R)+{k\over 2}(g_R^{-1}dg_R\w\vert \partial(g_R^{-1}dg_R))+\jp(u(dJ_R)\w\ve dJ_R)+$$
$$+ d(\mu\ve g_L^{-1}dg_L)-{k\over 2}(g_L^{-1}dg_L\w\vert \partial(g_L^{-1}dg_L))+\jp(u(dJ_L)\w\ve dJ_L),\eqno(45)$$
where
$$J_{L}=g_{L}\mu g_{L}^{-1}+k\partial_\si g_{L} g_{L}^{-1},$$
 $$J_{R}=-g_{R}\mu g_{R}^{-1}-k\partial_\si g_{R} g_{R}^{-1}.$$
 Before giving the interpretation of the reduced symplectic manifold in terms of the deformed
 gauged WZW model, let us first study the residual gauge symmetries of the form $\ti\om_u$.
 We recall that the residual gauge group $H$ is the normalizer of the Cartan algebra $\T_S$. We can make it 
 smaller by further gauge fixing. Thus we suppose that the variable $J_L^{\nu,0}(=-J_R^{\nu,0})$ 
 takes values only in the Weyl alcove of $\T_S$. (We remind that the Weyl alcove is the fundamental
 domain of the action of the affine Weyl group of $LS$ on $\T_S$). With this restriction 
 the residual gauge group   becomes just the Cartan torus $T_S$ acting
 as
 $$t_S\tr (g_L,g_R)=(t_Sg_L,t_Sg_R), \quad t_S\in T_S. \eqno(46)$$
Indeed, replacing $g_{L,R}$ by $t_Sg_{L,R}$ in (45), the form $\ti\om_u$ transforms as
$$\ti\om_u\to \ti\om_u+ d(J_L+J_R\ve t_S^{-1}dt_S)=\ti\om_u,$$
since the term $d(J_L+J_R\ve t_S^{-1}dt_S)$ vanishes due to the constraint $J_L^{\nu,0}+J_R^{\nu,0}=0$.
It is important to stress that the   parametrization (44) of the double $D$ via the variables $\mu,g_L,g_R$ gave
rise to another gauge symmetry of the form $\ti\om_u$ which is related to the ambiguity
of the chiral decomposition (44).  Indeed, if we pick arbitrary element $t_K$ from the Cartan torus
$T_K$ then it holds  
 $$(\chi, g)= \k(0,g_L)(\mu,e_{LK})(0,g_R)^{-1}=  \k(0,g_Lt_K)(\mu,e_{LK})(0,g_Rt_K)^{-1}.$$
 This means that the full residual gauge group of the form $\ti\om_u$ is $T_S\times T_K$ acting
 as
 $$(t_S,t_K)\tr (g_{L},g_{R})=(t_Sg_{L}t_K,t_Sg_Rt_K),\quad t_S\in T_S,\ t_K\in T_K.$$
 The reader may find strange that we have somewhat artificially augmented the residual 
 gauge symmetry of the Semenov-Tian-Shansky form $\om_u$ by expressing it in the new ambiguous variables $\mu,g_L,g_R$.
 However, the benefit of this parametrization consists in the fact that  in the  form $\ti\om_u$  
 the  variables $g_L$ and $g_R$ get disentangled.   The form $\ti\om_u$ is defined on the manifold
 $LK \times \A_K\times LK$ and its pull-back on $D$ via the map (44) gives the Semenov-Tian-Shansky
 form $\om_u$.  Obviously, it holds $D=(LK \times \A_K\times LK )/ T_K$.    We conclude this section
 by an observation, that the Hamiltonian (31)  of the $u$-WZW model descends to the reduced Hamiltonian
 $H_r$ (cf. the general discussion in Section 4.2).  Thus our symplectic reduction has produced 
 a new dynamical system $(M_r,\om_r, H_r)$ that will be interpreted in the next subsection as
 the deformed gauged WZW model.

 \subsection{Interpretation}

\noindent The gauged WZW model is  a dynamical system and its symplectic structure has been
thoroughly investigated e.g. in Sec. 3.2 and in Appendix A of  \cite{Gaw3}.  We report  here Gaw\c edzki's results
 in the
language of the left-right movers, by considering  maps $m_L,m_R:\br\to K$ fulfilling
$$ (\partial_\xi m_{L,R}m^{-1}_{L,R},\S)_\K=0,\eqno(47a)$$
$$m_{L,R}(\xi+2\pi)=e^{-{2\pi\nu\over k}}m_{L,R}(\xi)e^{{2\pi\mu\over k}}, \eqno(47b)$$
where $\mu$ is in the Weyl alcove of $\T_K$ and $\nu$ in the Weyl alcove of $\T_S$. The symplectic
form of the gauged WZW model is then given by the following expression (cf. Eq. (A.1) of \cite{Gaw3})
$$\om^{K/S}=- {k\over 2}(m_L^{-1}dm_L\w\ve \partial_\xi(m_L^{-1}dm_L))    +{k\over 2}(m_R^{-1}dm_R\w\ve \partial_\xi(m_R^{-1}dm_R))        $$
$$- {1\over 2}((m_L^{-1}dm_L)(0)-m_L(0)^{-1}{2\pi d\nu\over k}m_L(0), \w d\mu)_\K  -{1\over 2}((dm_L m_L^{-1})(0),\w d\nu)_\K$$
 $$+ {1\over 2}((m_R^{-1}dm_R)(0)-m_R(0)^{-1}{2\pi d\nu\over k} m_R(0), \w d\mu)_\K  +{1\over 2}((dm_R m_R^{-1})(0),\w d\nu)_\K.$$
In writing the form $\om^{K/S}$, we have switched from Gaw\c edzki's notations to ours (e.g. we have used
$(.,.)_\K$  instead of Tr$(.,.)$ etc.), nevertheless $\om^{K/S}$  still does not quite resemble our reduced form $\ti\om_{u=0}$.
In fact, we should note that Gawedzki's  chiral movers are quasiperiodic (cf. (47b)) while we use the periodic 
fields $g_{L,R}(\si)$. Indeed, if we perform  a transformation 
$$m_{L,R}(\xi)=e^{- {\nu\xi\over k} }g_{L,R}(\xi) e^{{\mu\xi\over k}},$$
the conditions (47)  become
$$ (g_{L,R}\mu g_{L,R}^{-1}+k\partial_\si g_{L,R} g_{L,R}^{-1} -\nu,\S)_\K=0,\eqno(48a)$$
$$g_{L,R}(\xi+2\pi)=g_{L,R}(\xi)\eqno(48b)$$
and the form $\om^{K/S}$ transforms to
$$\om^{K/S}=  d(\mu\ve g_L^{-1}dg_L-g_R^{-1}dg_R)-{k\over 2}(g_L^{-1}dg_L\w\vert \partial(g_L^{-1}dg_L))+{k\over 2}(g_R^{-1}dg_R\w\vert \partial(g_R^{-1}dg_R)).\eqno(49)$$
It is not difficult to find out that  the form (49) coincides with  the form $\ti\om_{u=0}$ given by (45) and
the constraints (48a) are, respectively, the constraints (40).
\vskip1pc
\noindent We observe that the symplectic reduction of the $u$-WZW model  for $U=0$ gives the
standard gauged WZW model. Therefore, if we switch on a non-trivial $U$, we  interpret the reduced
theory as the $u$-deformed gauged WZW model.   
\section{Conclusions and outlook}
In the present paper, we have presented a thorough discussion of the gauging of the deformed
WZW models. After the general derivation of 
the  quasi-adjoint actions (20a) and (20b), which are to be  gauged in general case, we have worked out
the $u$-deformed WZW model as an example.  Moreover, in Sections 3.2 and 3.3, we have
also introduced  the moment maps $\rho\circ \L_{L,R}$ which can be used for deforming
the procedure of the null gauging of the WZW models   \cite{5,KT} .    
\vskip1pc
\noindent  The main  open issue concerning the deformed WZW models is a quantization.
Since we dispose of the rather explicit description of the Poisson brackets of the deformed
WZW models (cf. Section 4.1) it seems to be doable to identify the operator algebra
of the quantum deformed model and also  the unitary representations of this algebra.  What seems to be
  more difficult, however,  is to extract  from the deformed WZW  theories general axioms of the deformed vertex
algebras. We find this  problem exciting and we wish to deal with it in future.


\begin{thebibliography}{10}
 

 
 \bibitem{AM}{A.Yu. Alekseev and A.Z. Malkin, Symplectic structures associated to Lie-Poisson 
groups, {\it Commun. Math. Phys.} {\bf 162}  (1994)  147-174, [hep-th/9303038]}
\bibitem {Mad}{J. Balog, L. Feh\'er and L. Palla, Chiral Extensions of the WZNW Phase Space, Poisson-Lie Symmetries and  Groupoids, {\it Nucl.Phys.} {\bf B568} (2000) 503-542 , [hep-th/9910046]}
\bibitem{FR}{H. Flaschka and T. Ratiu, Convexity theorem for Poisson actions of compact Lie groups, {\it Ann. Sci. Ecole Norm. Sup.}  {\bf 29} (1996) 787-809}
 \bi{5}{P. Forg\'acs, A. Wipf, J. Balog, L. Feh\'er and L. O'Raifeartaigh,  Liouville and Toda theories
as conformally reduced WZNW theories, {\it Phys. Lett.} {\bf B227} (1989) 214-220}
\bi{Gaw}{K. Gaw\c edzki,   Classical origin of quantum group symmetries in WZW conformal field theory,
{\it Commun. Math. Phys.} {\bf 139} (1991) 201-213}
\bi{Gaw2}{K. Gaw\c edzki, Topological actions in two-dimensional quantum field theories, In: {\it Non-perturbative quantum field theory, eds. G. Õt Hooft, A. Jaffe, G. Mack, P. K. Mitter, R. Stora} (Plenum Press, New York,1988)   p.101-141}
\bi{Gaw3}{K. Gaw\c edzki,  Boundary WZW, G/H, G/G and CS theories, {\it Annales Henri Poincar\'e} {\bf 3} (2002) 847-881, [hep-th/0108044]}
\bi{Kas}{C. Kassel, {\it Quantum Groups},  Springer -Verlag 1995}
\bibitem{K}{C. Klim\v c\'\i k,  Quasitriangular WZW model, {\it Rev. Math. Phys.} {\bf 16} (2004) 679-808,  [hep-th/0103118]}
\bi{K1}{C. Klim\v c\'\i k, Quasitriangular chiral WZW model in a nutshell, {\it Prog. Theor. Phys.Suppl}
 {\bf  144} (2001) 119-124, [hep-th/0108148]} 
\bi{K2}{C. Klim\v c\'\i k, Poisson-Lie symmetry and $q$-WZW model, to appear in the {\it Proceedings of the 4th International  Symposium On Quantum Theory And Symmetries (QTS-4), Varna Free University,  Bulgaria, 15-21 August 2005}, [hep-th/0511003]}
\bi{KS}{C. Klim\v c\'\i k and P. \v Severa, Open strings and D-branes in WZNW model,
{\it Nucl. Phys.} {\bf B488} (1997) 653-676, [hep-th/9609112]}
\bi{KT}{C. Klim\v c\'\i k and A.A. Tseytlin, Exact four-dimensional string solutions and Toda-like sigma models from 'null-gauged' WZNW theories, {\it Nucl. Phys.} {\bf B424} (1994) 71-96, [hep-th/9402120]}
\bi{OR}{J-P. Ortega, T. Ratiu, {\it Momentum maps and Hamiltonian reduction} (Birkhauser, Boston, 2004)}
\bi{ST1} {M. Semenov-Tian-Shansky, Dressing transformations and Poisson groups actions, {\it Publ.RIMS} {\bf 21}, 
Kyoto Univ. (1985) 1237-1260}
 \bi{ST2}{M. Semenov-Tian-Shansky,  Poisson Lie Groups, Quantum Duality Principle and the Twisted Quantum Double, 
{\it Theor. Math. Phys.}  {\bf 93} (1992) 1292-1307, [hep-th/9304042]}
 \bibitem{W} {E. Witten, Non-Abelian bosonisation in two dimensions,  {\it Commun. Math. Phys.} {\bf 92}  (1984) 455-472}
\bibitem{Wi3}{E. Witten, On holomorphic factorization of WZW and coset models, {\it Commun. Math. Phys.} {\bf 144} (1992) 189-212}
 

 

\end{thebibliography}
\end{document}